\documentclass[11pt]{article}
\usepackage{graphicx}  % standard LaTeX graphics tool

\usepackage{mathdots}

\usepackage{color}

\usepackage{a4}
\usepackage{amssymb,latexsym}%,amsmath}
\usepackage[mathscr]{eucal}
\usepackage{citesort}
\usepackage{url}
\usepackage{ntheorem}

\newcommand{\omicron}{o}%
\newcommand{\End}{{\mathrm{End}}}%
\definecolor{bluem}{rgb}{0,0,0.5}

\definecolor{mycolor}{cmyk}{0.5,0.1,0.5,0}
\definecolor{michel}{rgb}{0.5,0.9,0.9}
\definecolor{turquoise}{rgb}{0.25,0.8,0.7}
\definecolor{bluem}{rgb}{0,0,0.5}

\definecolor{MDB}{rgb}{0,0.08,0.45}
\definecolor{MyDarkBlue}{rgb}{0,0.08,0.45}

\definecolor{MLM}{cmyk}{0.1,0.8,0,0.1}
\definecolor{MyLightMagenta}{cmyk}{0.1,0.8,0,0.1}

\definecolor{HP}{rgb}{1,0.09,0.58}

\newcommand{\Ker}{{\mathrm{Ker}}}

\newcommand{\hhypo}{\,\,\widehat{\!\!\hyp}_i}
\newcommand{\Hnhypo}{\,\,\widehat{\!\!\hyp}}
\newcommand{\Hhypo}{\,\,\widehat{\!\!\hyp}_i}
\newcommand{\Sp}{S^+}
\newcommand{\Sm}{S^-}

\newcommand{\mTB}{m_{\mathrm{TB}}(\Sp)}
\newcommand{\mTBp}{m_{\mathrm{TB}}(\Spti)}
\newcommand{\mTBm}{m_{\mathrm{TB}}(\Smti)}
\newcommand{\trr}{{\tilde r}}
\newcommand{\fourg}{\,{}^{4}g}%
\newcommand{\threeg}{h}%
\newcommand{\tn}{{\tilde n}}
\newcommand{\tm}{{\tilde m}}
\newcommand{\tD}{{\tilde D}}
\newcommand{\tdelta}{{\tilde \delta}}

\newcommand{\tpi}{{\tilde \pi}}
\newcommand{\tmu}{{\tilde \mu}}
\newcommand{\tnu}{{\tilde \nu}}
\newcommand{\trho}{{\tilde \rho}}
\newcommand{\tgamma}{{\tilde \gamma}}
\newcommand{\tPhi}{{\tilde \Phi}}
\newcommand{\ttau}{{\tilde \tau}}
\newcommand{\talpha}{{\tilde \alpha}}
\newcommand{\tbeta}{{\tilde \beta}}

\newcommand{\tLambda}{{\tilde \Lambda}}

\newcommand{\FS}       %{F_1} %
                  {F}
\newcommand{\HS} %{F_2}
       {H_{\mbox{\scriptsize volume}}}

\newcommand{\eeal}[1]{\label{#1}\end{eqnarray}}
\newcommand{\C}{{\mathbb C}}%\newcommand{\C}{\mathbb C}%
\newcommand{\bed}{\begin{deqarr}}
\newcommand{\eed}{\end{deqarr}}
\newcommand{\bedl}[1]{\begin{deqarr}\label{#1}}
\newcommand{\eedl}[2]{\arrlabel{#1}\label{#2}\end{deqarr}}

\newcommand{\tg}{{\widetilde{g}}}
\newcommand{\tO}{{\widetilde{O}}}
\newcommand{\tiota}{{\widetilde{\iota}}}
\newcommand{\tom}{{\tilde{o}}}
\newcommand{\tnabla}{{\widetilde{\nabla}}}
\newcommand{\tI}{{\widetilde{I}}}
\newcommand{\tL}{{\widetilde{L}}}
\newcommand{\tN}{{\widetilde{N}}}
\newcommand{\mcO}{{\mycal O}}
\newcommand{\mcU}{{\mycal U}}

\newcommand{\mcN}{{\mycal N}}
\newcommand{\mcK}{{\mycal K}}

\newcommand{\bel}[1]{\begin{equation}\label{#1}}
\newcommand{\bea}{\begin{eqnarray}}
\newcommand{\bean}{\begin{eqnarray}\nonumber}
\newcommand{\bbean}{\begin{eqnarray*}}
\newcommand{\eean}{\end{eqnarray*}}
\newcommand{\beal}[1]{\begin{eqnarray}\label{#1}}
\newcommand{\eea}{\end{eqnarray}}
\newcommand{\Eq}[1]{Equation~\eq{#1}}

\newcommand{\qed}{\hfill $\Box$ \medskip}
\newcommand{\proof}{\noindent {\sc Proof:\ }}
\newcommand{\be}{\begin{equation}}
\newcommand{\eeq}{\end{equation}}
\newcommand{\ee}{\end{equation}}
\newcommand{\beqa}{\begin{eqnarray}}
\newcommand{\eeqa}{\end{eqnarray}}
\newcommand{\beqan}{\begin{eqnarray*}}
\newcommand{\eeqan}{\end{eqnarray*}}
\newcommand{\ba}{\begin{array}}
\newcommand{\ea}{\end{array}}

\newcommand{\Id}{\mbox{\rm Id}} %identity matrix
\newcommand{\hyp}{\mycal S}

\newcommand{\mcM}{{\mycal M}}
\newcommand{\mcD}{{\mycal D}}

\newcommand{\mcW}{{\mycal W}}
\newcommand{\mcV}{{\mycal V}}

\newtheorem{Theorem} {\sc  Theorem\rm} [section]
\newtheorem{Corollary} [Theorem] {\sc  Corollary\rm}
\newtheorem{Lemma} [Theorem] {\sc  Lemma\rm}
\newtheorem{Proposition} [Theorem] {\sc  Proposition\rm}

\theorembodyfont{\upshape}
\newtheorem{Remark}[Theorem]{\sc  Remark\rm}

\DeclareFontFamily{OT1}{rsfs}{} \DeclareFontShape{OT1}{rsfs}{m}{n}{
<-7> rsfs5 <7-10> rsfs7 <10-> rsfs10}{}
\DeclareMathAlphabet{\mycal}{OT1}{rsfs}{m}{n}
\def\scri{{\mycal I}}%
\def\scrip{\scri^{+}}%

{\catcode `\@=11 \global\let\AddToReset=\@addtoreset}
\AddToReset{equation}{section}

\newcounter{mnotecount}[section]

\renewcommand{\themnotecount}{\thesection.\arabic{mnotecount}}

\newcommand{\mnote}[1]%{}%
{\protect{\stepcounter{mnotecount}}$^{\mbox{\footnotesize
$%\!\!\!\!\!\!\,
\bullet$\themnotecount}}$ \marginpar{%\color{red}%
\raggedright\tiny\em $\!\!\!\!\!\!\,\bullet$\themnotecount: #1} }

\newcommand{\warn}[1]%{}
{\protect{\stepcounter{mnotecount}}$^{\mbox{\footnotesize
$%\!\!\!\!\!\!\,
\bullet$\themnotecount}}$ \marginpar{%\color{red}%
\raggedright\tiny\em $\!\!\!\!\!\!\,\bullet$\themnotecount: {\bf
Warning:} #1} }

\newcommand{\R}{\mathbb R}

\newcommand{\N}{\mathbb N}

\newcommand{\eq}[1]{(\ref{#1})}

\newcommand{\ext}{\mathrm{ext}}

\newcommand{\bmcO}{\,\,\overline{\!\!\mcO}}
\newcommand{\tmcO}{\,\,\,\widetilde{\!\!\!\mcO}}
\newcommand{\tmcM}{\,\,\,\,\,\widetilde{\!\!\!\!\!\mcM}}
\newcommand{\tMM}{\,\,\,\,\,\widetilde{\!\!\!\!\!\mcM}}%

\newcommand{\Vect}{\mathrm{Vect}}

\newcommand{\beqar}{\begin{deqarr}}
\newcommand{\eeqar}{\end{deqarr}}

\newcommand{\beaa}{\begin{eqnarray*}}
\newcommand{\eeaa}{\end{eqnarray*}}

\newcommand{\zg}{\mathring{g}}

\newcommand{\tl}{{\tilde \ell}}%
\newcommand{\tell}{{\tl}}%
\newcommand{\scrim}{{\scri^-}}%
\newcommand{\scripo}{{\scri^+_0}}%
\newcommand{\scripm}{{\scri^\pm}}%
\newcommand{\scripmo}{{\scri^\pm_0}}%

\newcommand{\beq}{\begin{equation}}

\newcommand{\zpsi}{{\mathring \psi}}%
\newcommand{\zhyp}{{\,\,\,\mathring{\!\!\! \hyp}}}%
\newcommand{\tM}{\widetilde M}
\newcommand{\zM}{\,\,\,\,\,\mathring{\!\!\!\!\!\mcM}}%
\newcommand{\tgf}{{^4\tilde g}}%
\newcommand{\tgfe}{{\tgf_\epsilon}}%
\newcommand{\tfourg}{{\tgf}}%
\newcommand{\hypo}{{\hypi}}%{{\hyp_{t_i}}}%
\newcommand{\hypi}{\hyp_i}%{{\hyp_{t_i}}}%
\newcommand{\mcnt}{\mcN}%
\newcommand{\Spto}{S^+_i}%{S^+_{t_i}}%
\newcommand{\Sptimo}{S^+_{{i-1}}}%
\newcommand{\Spti}{\Spto}%{S^+_{t_i}}%
\newcommand{\oSpti}{\mathring{S}^+_i}%{S^+_{t_i}}%
\newcommand{\Smti}{\Smto}%{S^+_{t_i}}%
\newcommand{\Sptio}{S^+_{{i_0}}}%
\newcommand{\myset}{\mcU}%
\newcommand{\Scrip}{\scrip}%
\newcommand{\io}{i_0}%
\newcommand{\dU}{{\,\,\,\dot{\!\!\!\mcU}}{}}%
\newcommand{\dD}{{\,\,\dot{\!\!\mcD}}{}}%

\begin{document}
\title{On Mason's rigidity theorem}

\author{Piotr T.\ Chru\'sciel\thanks{%Currently at the Albert Einstein Institute, Golm.
%Partially supported by a Polish
%Research Committee grant 2 P03B 073 24.
E-mail
    \protect\url{Piotr.Chrusciel@lmpt.univ-tours.fr}, URL
    \protect\url{ www.phys.univ-tours.fr/}$\sim$\protect\url{piotr}}
  \\ LMPT,
F\'ed\'eration Denis Poisson, Tours
\\
Mathematical Institute and Hertford College, Oxford\\
\\
\\
  Paul Tod\thanks{{ E--mail}: \protect\url{paul.tod@st-johns.ox.ac.uk}}
\\
Mathematical Institute and St John's College, Oxford}

\maketitle

\begin{abstract}
Following an argument proposed by Mason, we prove that there are no algebraically special
asymptotically simple vacuum space-times with a smooth, shear-free, geodesic congruence of
principal null directions extending transversally to a cross-section of $\scrip$. Our analysis leaves
the door open for escaping this conclusion if the congruence is not smooth, or not transverse to
$\scrip$. One of the elements of the proof is a new rigidity theorem for the Trautman-Bondi mass.
\end{abstract}

\tableofcontents

\section{Introduction}
\label{introduction}

It is a long standing conjecture that the only {vacuum}
algebraically special asymptotically simple space-time is Minkowski
space. Arguments towards a proof  have been presented
in~\cite{Mason:AlgSpec}. The aim of this work is to establish the
conjecture under a set of restrictive conditions, with the aid of a
rigidity theorem for Trautman-Bondi mass, a complete proof of which
has not been presented previously.

A space-time $(\mcM,g)$ is said to admit a conformal boundary
completion at  infinity if there exists a manifold $\tmcM=\mcM\cup
\scri$ with boundary $  \scri$ and a function $\Omega$ on $\tmcM$
vanishing precisely on $\scri$, with nowhere vanishing gradient
there, such that the metric $\tg:=\Omega^2 g$ extends smoothly to a
tensor field with Lorentzian signature  defined on $\tmcM$. We
denote by $\scrip$, respectively $\scrim$, the set of points on
$\scri$ which are end-points of null future directed, respectively
past directed, geodesics. We will
say that%
\footnote{Recall that a space-time is a time-oriented Lorentzian
manifold, the topology of which is assumed to be metrisable. In view
of our extensive use of the NP formalism the signature $(+---)$ is
used.}
$(\mcM,g)$ is \emph{past asymptotically simple} if every maximally
extended null geodesic acquires a past end-point on $\scri^-\subset
\scri$. \emph{Future asymptotic simplicity} is defined by changing
time-orientation. Following~\cite{NewmanAF}, we use the term
\emph{asymptotic simplicity}  if $\mcM$ does not contain closed
timelike curves, and if both past and future asymptotic simplicity
hold. An embedded submanifold of $\scri$ will be said to be a
\emph{cross-section} if it meets generators of $\scri$
transversally, and at most once each.

Asymptotically simple space-times with null conformal boundaries are
known to be globally hyperbolic~\cite{NewmanAF}, with contractible
Cauchy surfaces, with $\scrip$ and $\scrim$ containing $\R\times
S^2$, where the $\R$ factor corresponds to motions along the null
geodesic generators. Furthermore, $\scri$ reduces to two copies of
$\R\times S^2$ if one assumes that the extended space-time
$(\tmcM,\tg)$ is strongly causal at $\scri$.

It appears of some interest to consider algebraically special
space-times which are {asymptotically simple to the past},
without necessarily being asymptotically simple.%
\footnote{We are grateful to an anonymous referee for pointing out
this possibility to us.}
Such space-times  could describe e.g. the formation of a black hole
in a space-time without singularities in the past.

 Recall that a space-time
$(\mcM,g)$ is \emph{algebraically special} if at every point there
exists a null vector $\ell$  such that   the Weyl tensor $ C_{\mu\nu
\rho \sigma}$ satisfies
\bel{algspecSpt} C_{\mu\nu \rho [\sigma}\ell_{\pi]} \ell^\rho
\ell^\nu = 0
 \;.
\ee
Assume that $(\mcM,g)$ is vacuum, or that the Ricci tensor
satisfies a set of restrictions listed in detail below. Then, on
regions where the
$\ell$'s can be chosen to produce a smooth vector field, %in vacuum
%and in some restricted non-vacuum cases,
near those orbits along
which the Weyl tensor isn't zero everywhere,
$\ell$ can be rescaled%
\footnote{A priori this can be done only locally; however, in
globally hyperbolic space-times (which is the case here) this can
always be done globally when $\ell$ is globally smooth.}
so that its integral curves are null affinely parameterized
geodesics without shear. Conversely, %in vacuum, or under the
%restrictions on the Ricci tensor to which we return below,
the existence of such a congruence implies \eq{algspecSpt}.

  In
asymptotically simple space-times the integral curves of $\ell$
extend smoothly to the conformal boundary at their end points.
The question then arises, whether a suitable rescaling $\tl$ of
$\ell$ extends by continuity to a smooth vector field defined on the
set
\beaa
 \lefteqn{ \mcM\cup\underbrace{\{p\in \scri| \mbox{ $p$ is an end point
of
 precisely one integral
 curve of $\ell$}\}}_{\mcU}} &&
  \\
  &&\phantom{xxxxxxxxxxxxxxxxxxxxxxxxxxxxxxxxxxxxxx}\subset\tmcM:=\mcM\cup \scri
 \;.
\eeaa
%$$
%.
Since the integral curves of $\ell$ intersect $\scri$ transversally,
$\tl$ is transverse to $\mcU$. However, neither continuity nor
differentiability of $\tell$ at $\mcU$ are clear. Further, $\tl$
might become singular as the boundary $\overline\mcU\setminus \mcU$
is approached, or perhaps develop zeros there. Problems will clearly
arise at points at which more than one integral curve of $\ell$
meets $\scrip$, assuming that $\tl$ can be defined at those points
at all.
% A criterion for smoothness of $\tell$ at $\scri$ is given in
%Appendix~\ref{SPaul}.
% A criterion for smoothness of $\tell$ at $\scri$ is given in
One of our results here (see Appendix~\ref{SPaul}) is the proof that
smoothness and transversality to $\scrip$ of
$$
 \tell:=\Omega^{-2}\ell
$$
is equivalent to the non-existence of zeros of   the complex
divergence $\rho=\bar m^\mu m^\nu\nabla_\mu\ell_\nu$  {(see,
e.g.,~\cite{NewmanTod} for details of the definition of $m^\mu$)} of
the congruence defined by $\ell$ in a neighborhood of $\scrip$.

Assuming that  {$\ell$} is globally well defined, in
Section~\ref{Scss} we prove:

\begin{Theorem}
 \label{Tm1v0}
The following  {set of conditions is incompatible}  {in vacuum}:

\begin{enumerate}
\item
$(\mcM,g)$ is   {past asymptotically simple and contains a
contractible
Cauchy surface.}%
\footnote{\label{FNew}As already pointed out, all these conditions
will hold by~\cite{NewmanAF} if the space-time is asymptotically
simple.}

\item %There exists on $\tmcM$ a smooth, null, shear-free, geodesic
%vector field $\tell$ satisfying
%%
%\bel{algspecv} \tC_{\mu\nu \rho [\sigma}\tl_{\pi]} \tl^\rho \tl^\nu
%= 0
% \;,
%\ee
%%
%where $\tC_{\mu\nu\rho\sigma}$ is the Weyl tensor of $\tg$.
There exists on $\mcM$ a smooth, null, shear-free, geodesic vector
field $\ell$.
%, so that
%%
%\bel{algspecvn}
% C_{\mu\nu \rho [\sigma}\ell_{\pi]}  \ell^\rho
% \ell^\nu = 0
% \;,
%\ee
%%
%where $C_{\mu\nu\rho\sigma}$ is the Weyl tensor of $g$.
\item We have $\scrim\approx \R\times S^2$,
and there exists a compact cross-section
 $\Sp$ of  $ \Scrip$ near
which a rescaling $\tell$ of $\ell$ extends by continuity to a
smooth vector field which is transverse to $\Scrip$.
\end{enumerate}

The statement   remains true for non-vacuum metrics if the dominant
energy condition holds and if the Newman-Penrose components
$\Phi_{00}$, $\Phi_{01}$, $\Phi_{02}$ and $\Lambda$ of the  Ricci
tensor%
\footnote{\label{f3}Here and elsewhere, we follow the conventions of
\cite{NewmanTod} for the NP spin-coefficient formalism, so that
these conditions on the Ricci tensor are equivalent to the vanishing
of the scalar curvature and of $\Phi_{ABA'B'}o^Ao^B$, where $o^A$ is
the spinor obtained from $\ell$.}, associated to the congruence
defined by $\ell$, vanish, with
the remaining components decaying fast enough.%
\footnote{The exact decay rates needed can be found by chasing
through the calculations in~\cite{TafelBondi,NatorfTafel} that lead
to the Natorf-Tafel mass aspect formula \eq{Tafelform} below.}
\end{Theorem}

\begin{Remark}
Theorem~\ref{Tm1v0} holds under finite differentiability conditions
on $\tell$, but we have not attempted to determine the threshold; in
any case there are no \emph{a priori} reasons for a geodesic
shear-free null congruence to be globally $C^0$ or $C^1$, even if
the metric is smooth. Indeed, poorly differentiable examples can be
constructed in Minkowski space-time, see Section~\ref{Sndc} where
somewhat more general congruences are allowed.
\end{Remark}

\begin{Remark}
 \label{RGST}
As can be seen from footnote \ref{f3}, the Ricci tensor conditions
of Theorem~\ref{Tm1v0} will hold in electro-vacuum if $o^Ao^B
\varphi_{AB}=0$, where $\varphi$ is the Maxwell spinor, with $o^A$
as in Section~\ref{Smfasvs}.
\end{Remark}

{Theorem~\ref{Tm1v0} is similar in spirit to the results of
Mason~\cite{Mason:AlgSpec}.} The differences between our hypotheses
and those of~\cite{Mason:AlgSpec} are as follows: First,  algebraic
speciality  does \emph{not}  imply the
 \emph{smoothness} of  either  $\tl$ or $\ell$.
Next,  {neither existence nor transversality of $\tl$ at $\scrip$
are assumed} in~\cite{Mason:AlgSpec}. We further note that the
hypotheses of Theorem~\ref{Tm1v0} enforce non-vanishing of twist
throughout a region of $\mcM$ relevant for the proof (see
Proposition~\ref{Pnotwist} below), while more general configurations
are \emph{a priori} allowed in~\cite{Mason:AlgSpec}.%
\footnote{Theorem~\ref{Tnew} below allows configurations somewhat
more general than Theorem~\ref{Tm1v0}, but those are still less
general than indicated in~\cite{Mason:AlgSpec}.} Finally, our
argument requires the topology of $\scrim$ to be $\R\times S^2$
which, {for asymptotically simple space-times as considered
in~\cite{Mason:AlgSpec}}, is only known to be true~\cite{NewmanAF}
when $\mcM\cup \scrim$ is strongly causal.

The key idea stems from~\cite{Mason:AlgSpec}, but some steps of the
argument require careful reorganizations. The proof can be
structured as follows: We start, in Section~\ref{Smfasvs}, by
introducing a coordinate system based on the members of the
congruence. This allows us to construct a cut $\Sm $ of $\scrim$,
associated to the cut  $\Sp $ of $\scrip$, on which $\tell$ is
transverse. The calculations in~\cite{Mason:AlgSpec} subsequently
show that  the Trautman-Bondi mass  $\mTB $ of  $\Sp $  is the
negative  of that of $\Sm$.

One then wishes to appeal to the positive energy theorem to show
flatness of the metric near a cross-section of $\scrip$. This
requires controlled spacelike hypersurfaces, say $\hyp $, which are
constructed at the beginning of Section~\ref{Scss}. So, the positive
energy theorem of~\cite{CJL} implies that   $\mTB $ vanishes, and
that   $\hyp $ carries a timelike KID. An analysis of Killing
developments allows one to conclude that the initial data on the
$\hyp $  can be realized by embedding in Minkowski space-time; this
is in fact a new rigidity result for the Trautman-Bondi mass, see
Theorem~\ref{Tpos}.  {One concludes by showing that no congruences
with the properties listed exist near a Minkowskian $\scrip$.}

\medskip

We shall say that an algebraically special space-time is
\emph{non-branching} if it is either type $II$ {or $D$ }
everywhere\footnote{We allow the metric to be $I\!I$ at some
places and $D$ at others.}, or type $III$ everywhere, or type
$N$ everywhere. The point is that in these cases the Weyl
tensor does not allow branching of the principal null
directions. We then have the following related statement:

\begin{Theorem}
 \label{Tm2}
 The following conditions are incompatible:
 \begin{enumerate}
 \item $(\mcM,g)$ is  past
 asymptotically simple and contains a contractible
 Cauchy surface.$^{\mbox{\scriptsize\ref{FNew}}}$

 \item
$(\mcM,g)$ is non-branching, vacuum, with $  \scrim\approx \R\times
S^2$, and the complex divergence $\rho$ of the congruence has no
zeros near a compact cross-section $\Sp$ of $\scrip$.
 \end{enumerate}

The conclusion remains true for non-vacuum space-times if the
conditions on the Ricci tensor spelled out in Theorem~\ref{Tm1v0}
are met.
\end{Theorem}

Indeed,  assume that such a space-time exists. We show in
Appendix~\ref{Asmo} that $\ell$ can be chosen to be smooth
throughout $\mcM$, and in Appendix~\ref{SPaul} that
$\Omega^{-2}\ell$  is smooth and transverse at $S$. By
Proposition~\ref{Pflatr} below  $(\mcM,g)$ contains a flat region,
thus is of type $O$ there, which gives a contradiction.

\section{Non-differentiable congruences}
 \label{Sndc}

It appears of interest to find a set of hypotheses, alternative to
those of Theorem~\ref{Tm1v0},  which are compatible with at least
one space-time. A possible direction of enquiries is to admit
smoothness and transversality of $\tell$ near one or more sections
of $\scrip$, but allow singularities of $\tell$ in the space-time.
(The question of non-transversal congruences will be discussed in
Section~\ref{Sconcl}.) Now, consider any maximally extended null
geodesic $\gamma$ initially tangent to $\tell$ near $\scrip$. In the
argument below it is \emph{necessary} that the tangent to $\gamma$
remains proportional to $\tell$. If $\tell$ is allowed to become
singular, this last property might not hold, and it is easy to
construct congruences where this occurs. (Consider, for example, any
timelike curve $\Gamma$ in Minkowski space-time extending from $i^-$
to $i^+$, let $u$ denote the retarded time function based on
$\Gamma$, and let $\ell=du$ on $\mcM\setminus \Gamma$. Then every
integral curve of $\ell$, when followed from $\scrip$ towards
space-time, stops at $\Gamma$.) Clearly, any argument in which null
geodesics {need to be} followed from $\scrip$ to $\scrim$ has no
chance of succeeding in such situations.

In this last example, of a congruence based on a curve
$\Gamma$, one can  smoothly flow the geodesics through
$\Gamma$, landing on a second congruence generated by the past
light-cones issued from $\Gamma$. To accommodate such
situations in any kind of generality
would require considering multiple-valued congruences.%
\footnote{The density condition~\eq{newcond} below essentially
forbids multiple-valued congruences,  {which therefore appear to be
intractable by our arguments}.}

One could, however, enquire what happens if $\tell$ is smooth on a
dense set, and if one further assumes that null geodesics somewhere
tangent to $\tell$ remain tangent to $\tell$ at all those points at
which $\tell$ is defined.  The apparent difficulty of flowing along
a singular vector field $\tell$ is easily resolved by flowing along
the associated geodesics. Anticipating, in such situations
Proposition~\ref{Pnotwist} below does not hold anymore, and one
faces the problem of understanding what happens  along those
geodesics on which the twist vanishes. The hypothesis that the
space-time is smooth together with the Newman-Penrose equations
leads  then to the vanishing of some components $\psi_i$ of the Weyl
tensor along such geodesics, but the implications of this are not
clear. It is conceivable that this might again lead to a mass
changing sign as in Proposition~\ref{Pmasschange} below, which would
allow one to conclude, but this remains to be seen.

In spite of the above, some degree of singularity of $\ell$ can be
allowed, as follows. To obtain more control of the space-time  we
will assume full asymptotic simplicity,
and consider a sequence of spherical cuts $\Spti$ of $\scrip$ near
which $\tell$ is again assumed to be smooth and transverse. Let us
set
\beaa
 \oSpti& :=& \{p\in \Spti: \ \mbox{ $\ell$ is smooth in an $\mcM$-neighborhood
 of the   maximally }
  \\
  &&
  \mbox{extended null geodesic with tangent $\tell$ at its
 end point $p$}\}
 \;.
\eeaa
Rather than assuming that $\ell$ is smooth throughout $\mcM$, so
that $\oSpti=\Spti$, suppose instead that
\bel{newcond}
%$$
 \mbox{ $\oSpti$ is dense within $\Spti$}
  \;.
\ee

Let $\scri_0\subset \scri$ be defined as
$$
 \scri_0:= \{ p\in \scri | \ \mbox{ strong causality holds at $p$}
 \}
 \;,
$$
with $\scripmo=\scri_0\cap \scripm$.  According to
Newman~\cite{NewmanAF}, in asymptotically simple space-times
each of $\scripmo$ is diffeomorphic to $\R\times S^2$, with the
generators of $\scri$ tangent to the $\R $ factor, which we
parameterize by $u$; we choose $ u$ to be increasing to the
future. We let $\Spti$ be any $S^2$ included in $\scripo$ that
intersects every generator of $\scripo$ precisely once; such
sets will be called cross-sections of $\scrip$. (It actually
follows from Theorem~\ref{Tnew}, which we are about to state,
that $\scri_0=\scri$ under the hypotheses there.) We claim
that:

\begin{Theorem}
 \label{Tnew}
Let $(\mcM,g)$ be an asymptotically simple space-time with smooth
null asymptote $(\tmcM,\tg)$ such that  {$\scrim\approx \R\times
S^2$}. Assume that there exists a sequence of  cross-sections
$\Spti$ of $\scripo$, $i\in \N$, such that
$$
\scripo\subset\cup_{i\in\N}J^+(\Spti,\tMM)
 $$
together with  a  geodesic, shear free,
null vector field $\tell$ defined on a neighborhood of
$$
 \mcM\cup_{i\in\N} \Spti
% \;,
$$
%,
satisfying   \eq{newcond}.
% together with
%%
%\bel{algspecvo} \tC_{\mu\nu \rho [\sigma}\tl_{\pi]} \tl^\rho \tl^\nu
%= 0
% \;
%\ee
%%
%where $\tC_{\mu\nu \rho \sigma}$ is the Weyl tensor of $\tg$.
Assume that the dominant energy condition holds and that the
Newman-Penrose components $\Phi_{00}$, $\Phi_{01}$, $\Phi_{02}$ and
$\Lambda$ of the  Ricci tensor, associated to the congruence defined
by $\ell$, vanish, while the remaining ones decay fast enough. If
$\tl$ is transverse to $\cup_{i\in\N} \Spti$, then $( \mcM,g)$ is
the Minkowski space-time $\R^{1,3}$.
\end{Theorem}

\begin{Remark}
The Kerr congruence in Minkowski space-time (see, e.g., the appendix
to~\cite{LindNewman}) satisfies the hypotheses of
Theorem~\ref{Tnew}.
\end{Remark}

The proof of Theorem~\ref{Tnew} can be found at the end of
Section~\ref{Scss}.

 \medskip

\section{The metric form of algebraically-special vacuum solutions}
 \label{Smfasvs}

We now run through the derivation of the metric form of
algebraically-special space-times, first vacuum and then noting
the changes for non-vacuum. References for this
are~\cite{Mason:AlgSpec,Exactsolutions2,LindNewman}. The
general technique is to construct a coordinate system with the
aid of the geodesic and shear-free congruence and solve enough
of the Newman-Penrose spin-coefficient equations to obtain the
radial dependence of the metric. We follow~\cite{NewmanTod} for
the Newman-Penrose spin-coefficient equations, rather than the
version in \cite{Exactsolutions2}. We modify the derivations in
\cite{Exactsolutions2} and \cite{Mason:AlgSpec} in order to
connect the coordinate system to standard coordinates on
$\scrip$ from the start of the calculation.

% \ptc{dangerous renew commands for $\Spti$, $\hypo$, $\Spto$, $\scripo$ and $\Smti$, reset later down the text}
\renewcommand{\hhypo}{\Hnhypo}
\renewcommand{\Spti}{\Sp}
\renewcommand{\hypo}{\hyp }
\renewcommand{\Smti}{\Sm}
\renewcommand{\scripo}{\scrip}
\renewcommand{\Spto}{\Sp }
Consider a manifold $\Sp$ transverse to the generators of $\scrip$.
We use a local coordinate $u$ along the generators  and a local
complex coordinate $\zeta$ on $S^+$, so that the (degenerate) metric
of $\scripo$ is
$$-4\frac{d\zeta d\overline{\zeta}}{P^2}
\;.
$$
As the calculations that follow are purely local in $u$ and $\zeta$
we can, without loss of generality, choose $P=1+\zeta
\overline{\zeta}$. (If $S^+$ is a sphere, then $(u, \zeta,
\overline{\zeta})$ are Bondi coordinates at $\scripo$.) Recall that,
in $(\tmcM,\tg)$, the usual spinor dyad $(\tO^A,\tI^A)$ and
corresponding NP tetrad $(\tL^a,\tN^a,\tM^a,\overline{\tM}{}^a)$ are
related to the coordinates by
\begin{eqnarray*}
 \tN^a\partial_a=\partial_u\;, \\
 \tM^a\partial_a=\frac{P}{\sqrt{2}}\partial_{\zeta}\;,
\end{eqnarray*}
and
\begin{eqnarray*}
 \tN_adx^a=-d\Omega\;, \\
 \tM_adx^a=-\frac{\sqrt{2}}{P}d\overline{\zeta} \;,
\end{eqnarray*}
where the last equation is understood as pulled back to $\scripo$,
where $d\Omega$ pulls back to zero (the point is that there are
different ways of extending the coordinates into the interior).

By assumption,
%, we have (\ref{algspecSpt}) for a smooth vector field
%$\ell$. As already pointed out, by the Goldberg-Sachs Theorem in
%vacuum (see e.g.~\cite{Exactsolutions2}),
$\ell$ generates a
geodesic and shear-free null congruence and we may scale $\ell$ so
that it is affinely parameterized.  Therefore $\ell$ defines a
smooth spinor field $o^A$, which in turn can be scaled to be
parallelly-propagated along the congruence. In the NP formalism,
this is
\beq\label{o2} Do^A:=\ell^b\nabla_bo^A=0.\eeq
There remains a residual freedom to rescale $o^A$ by a nowhere-zero
function $F^0$ which is constant along the congruence, when $\ell$
rescales with $|F^0|^2$.

Under conformal-rescaling of the metric $\tg=\Omega^2g$, the
rescaling $\tell=\Omega^{-2}\ell$ takes the affinely normalized
geodesic vector field $\ell$ to an affinely normalized geodesic one,
leading to a vector field $\tell$ which is continuous at $\scri$ by
hypothesis.
The rescaling $\tom^A=\Omega^{-1}o^A$ likewise extends to $\scripo$,
where
\bel{aboveo1} \tom^A=B\tO^A+C\tI^A
 \ee
for smooth functions $B$ and $C$ on $\scripo$. The  assumption that
$\tell$ is transverse to $\scripo$on $\Spti$ translates to the
requirement that $B$ be nonzero on that part of $\scripo$. We extend
$B$ and $C$ into the interior as functions constant along the
congruence and then rescale $\tom^A$ to set $B=1$. Now, at
$\scripo$, we have
\beq \label{o1} \tom^A=\tO^A+L(u,\zeta, \overline{\zeta})\tI^A, \eeq
in terms of a function $L$ on $\scripo$. The assumption of
transversality implies that $L$ is a smooth function on $\Spti$. (We
could define $L$ independently of the scaling of $\tom^A$ as
$L=\tO_A\tom^A/\tI^B\tom_B$.)

Equation (\ref{o2}) implies that the spin-coefficients $\kappa$ and
$\epsilon$ are zero and, by assumption, $\sigma$ is also zero.

We extend the coordinates $u$ and $\zeta$ into the interior by
taking them to be constant along the geodesics of the congruence, so
that
\[Du=0=D\zeta\;,\]
and then $D=\partial/\partial r$ with $r$ as before. This fixes $r$
uniquely up to a shift of origin on each geodesic of the congruence.
A convenient (and standard) way to choose the origin in $r$ is next
to solve one of the spin-coefficient equations, (A.3a) as given in
\cite{NewmanTod} which, with the restrictions that we currently have
on the spin-coefficients, is just:
$$D\rho=\rho^2.$$
The solution of this is either $\rho\equiv 0$  or
\beq\label{r1} \rho=-(r+r^0+i\Sigma)^{-1} \eeq
where $r^0$ and $\Sigma$ are real functions of integration,
constant along the congruence (and so are functions only of
$(u,\zeta, \overline{\zeta})$; as before, we use the
superscript 0 for functions independent of $r$ but, as is
conventional, omit it from $\Sigma$). We choose the origin of
$r$ so that $r^0=0$. Note that, if $\rho \not \equiv 0$ and
$\Sigma$ ever vanishes, so that the twist of the congruence
vanishes, then $\rho$ is real and in this case $\rho$ will
diverge at a finite $r$. This is incompatible with smoothness
of the congruence,  unless $\rho$ identically vanishes, leading
to:

\begin{Proposition}
\label{Pnotwist} Under the hypotheses of Theorem~\ref{Tm1v0}, the
divergence $\rho$
 and the twist $\Sigma$ are nowhere vanishing on those integral
curves of $\ell$ which have end points on $\Sp$.
\end{Proposition}

\proof Since    $\tell $ is smooth, transverse, and geodesic, we
must have $\tell = \chi \Omega^{-2} \ell$ for some smooth nowhere
vanishing function $\chi$. We might therefore without loss of
generality assume, rescaling $\tell$ if necessary, that $\chi=1$.
From Equations~\eq{abovesc2} and \eq{sc2} of Appendix~\ref{SPaul} we
obtain
$$
 \rho =\Omega^2 \tilde \rho + \frac {D\Omega} \Omega= - \frac 1 r +
 O(r^{-2})
\;.
$$
where $\trho$ is associated with $\tell$ just as $\rho$ is
associated with $\ell$ (see Appendix~\ref{SPaul} for the
details of this). We conclude that $\rho$ has no zeros near
$\Sp$. Hence $\rho$ is not identically zero on any  of the
relevant members of the congruence, and since $\ell$ is smooth
by hypothesis,  \eq{r1} excludes zeros of $\Sigma$. {}
 \hfill\qed

Following Mason~\cite{Mason:AlgSpec}, who in turn follows Debney et
al.~\cite{DebneyKerrSchild}, we next consider the complex vector
field
$$W_a=o^B\nabla_a o_B.$$
By the geodesic, shear-free condition this is of the form
$o_A\tau_{A'}$ for some spinor field $\tau_{A'}$ with
$\overline{o}^{A'}\tau_{A'}=\rho$. By smoothness of the congruence,
$\rho$ is smooth in the interior, and   it does not vanish by
Proposition~\ref{Pnotwist}. Therefore we can define the spinor field
$\iota^A$, which makes up the NP dyad with $o^A$, via its complex
conjugate by $\overline{\iota}^{A'}=-\rho^{-1}\tau^{A'}$. Then
\beq \label{w1} W_a=-\rho o_A\overline{\iota}_{A'}=-\rho m_a, \eeq
and the spin-coefficient $\tau$ is also zero.

With the spin-coefficient equations numbered as in~\cite{NewmanTod},
from (A.3c) and (A.3p) with what we have now and the vacuum
equations we find that $\pi$ and $\lambda$ vanish. With the aid of
(A.3a), we calculate the exterior derivative of $W_a$ from
(\ref{w1}) as
\beq\label{w2}\nabla_{[a}W_{b]}=X_1\ell_{[a}m_{b]}+X_2\overline{m}_{[a}m_{b]},\eeq
in terms of two functions $X_1$ and $X_2$ whose precise form does
not concern us. Thus, in the language of differential forms,
$W\wedge dW=0$. We note the following, presumably
well known, complex version of the Frobenius theorem:

\begin{Lemma}
 \label{LFrob}
 There exist, locally, complex-valued functions $X_3$ and $X_4$ such
that
$$
 W=X_3dX_4 \;.
$$
\end{Lemma}

\proof Note that
\bel{novanish} W\wedge
\overline{W}=\rho\overline{\rho}m\wedge\overline{m}\neq 0
 \ee
by Proposition~\ref{Pnotwist}, which shows that the real and the
imaginary part of $W$ are nowhere vanishing, linearly independent.
Elementary algebra gives  $dW=W\wedge Z$ for some complex-valued
one-form $Z$.
The usual calculation shows that the two-dimensional distribution
defined by the collection of vector fields
$$
\{ X\in \Gamma TM:\quad W(X)=0
 \}
 $$
is integrable. Hence there exist, locally, complex functions
$\alpha$ and $\beta$, as well as real valued functions $f$ and $g$
such that
$$
W = \alpha df + \beta dg
 \;.
$$
\Eq{novanish} shows that $\alpha$ and $\beta$ are nowhere-vanishing,
and that $df$ and $dg$ are linearly independent. The equation
$W\wedge dW=0$ implies $\alpha/\beta=\varphi+i\psi$ for some
functions $\varphi=\varphi(f,g)$ and $\psi=\psi(f,g)$, hence
$$ W= \beta (\varphi df + dg + i \psi dg) \;.
$$
Consider the two-dimensional Riemannian metric
$$
 b:= (\varphi df + dg)^2 +   \psi^2 dg^2
 \;.
$$
By the uniformization theorem there exist, again locally, smooth
functions $x$, $y$ and $h$ such that
$$
 b = e^{2h}\Big( (dx)^2 + (dy)^2
 \Big) \;.
$$
Changing $y$ to $-y$ if necessary, at each point the
$b$-ON coframes $\{\varphi df + dg , \psi dg\} $ and $\{e^{h}
dx,e^{h}dy\}$ are rotated with respect to each other,
hence there exists a function $\theta=\theta(x,y)$ such that
$$
(\varphi df + dg + i \psi dg) = e^{h+i\theta}(dx+i dy)
 \;.
$$
The functions $X_3= \beta e^{h+i\theta}$ and $X_4=x+iy$ satisfy our claim.
 \qed

 \medskip

Returning to the problem at hand, either \eq{novanish}, or the
argument in the proof of Lemma~\ref{LFrob}, shows that the real and
imaginary parts of $X_4$ are independent functions.

By (\ref{w1}) and (\ref{w2}), both $dX_3$ and $dX_4$ are orthogonal
to $\ell$, so that both are functions only of $(u, \zeta,
\overline{\zeta})$, and we can determine them by looking at the
value of $W$ at $\scripo$. We have
\begin{eqnarray*}
W_adx^a&=&o^B\nabla_a o_Bdx^a\\
&=&\Omega\tom^B(\tnabla_{AA'} \tom_B+\Upsilon_{BA'}\tom_A)dx^a,
\end{eqnarray*}
where $\Upsilon_a=\partial_a\log\Omega$, we use the rules for
conformal transformation of the spinor connection
given in~\cite{PenroseRindlerv12} and we use the rescaled dyad of (\ref{resc}). We calculate this from (\ref{o1}) and
pullback to $\scripo$ to find that, at $\scripo$,
\beq \label{w3}
W_adx^a=\tM_adx^a=-\frac{\sqrt{2}}{P}d\overline{\zeta}. \eeq
However, from what we have said above about $X_3$ and $X_4$,
(\ref{w3}) holds everywhere, so that, by (\ref{w1}), in the interior
\beq \label{m1} m_adx^a=\frac{\sqrt{2}}{\rho P}d\overline{\zeta}.
\eeq
It follows at once from this that, in terms of the NP operators
$\Delta$ and $\delta$,
$$\Delta\zeta=0=\delta\overline\zeta,$$
while $%\overline
{\delta}\zeta=-\frac{\overline{\rho} P}{\sqrt{2}}$.

We need covariant and contravariant expressions for the rest of the
NP tetrad. We have
\begin{eqnarray*} \Delta &=&(\Delta u)\partial_u+(\Delta
r)\partial_r,\\
\delta &=&(\delta u)\partial_u+(\delta
r)\partial_r-\frac{\overline{\rho} P}{\sqrt{2}}\partial_{{\zeta}}.
\end{eqnarray*}
From the commutator $[\Delta, D]$ (given in~\cite{NewmanTod})
$$D\Delta u=0$$
so that $\Delta u=X_6(u, \zeta, \overline{\zeta})$ for some function
$X_6$, and analogously, from the commutator $[\delta, D]$ (using
(\ref{r1})) $\delta u=\overline{\rho}X_7$ in terms of another
function $X_7$ of $(u, \zeta, \overline{\zeta})$.
Since $\ell^a$ is null, we have
$$\ell_adx^a=Adu+Bd\zeta+\overline{B}d\overline{\zeta}$$
for some real $A$ and complex $B$ (not to be confused with $B$
appearing in \eq{aboveo1}), and then normalization against $m^a$ and
$n^a$ forces
$$
AX_6=1\;,\quad
A\overline{\rho}X_7-{B}\frac{\overline{\rho}P}{\sqrt{2}}=0\;,$$
so that $A$ and $B$ are independent of $r$ and can be found
from $\tell$ at $\scripo$. There we have (\ref{o1}) so that, on
$\scripo$,
\begin{eqnarray}
\ell_adx^a&=&(\tL_a+L\overline{\tM}_a+\overline{L}\tM_a)dx^a\nonumber\\
&=&du-\frac{\sqrt{2}L}{P}d\zeta-\frac{\sqrt{2}\overline{L}}{P}d\overline{\zeta}\label{l1}.
\end{eqnarray}
Now we argue as for $m_adx^a$: from what we have deduced already for
$\ell_adx^a$, we know that (\ref{l1}) holds in the interior. This
gives the NP tetrad in the covariant form as
\bea D&=&\partial_r\label{l2}\\
\Delta&=&\partial_u+H\partial_r\label{n2}\\
\delta&=&-\frac{\overline{\rho}P}{\sqrt{2}}(\partial_{%\overline
{\zeta}}
+\frac{\sqrt{2}%\overline
{L}}{%\overline{\rho}
P}\partial_u-Q\partial_r)\label{m2}
\eea
where $H$ and $Q$ are still to be determined, and in the
contravariant form as
\bea \ell_adx^a&=&du-\frac{\sqrt{2}L}{P}d\zeta
-\frac{\sqrt{2}\overline{L}}{P}d\overline{\zeta}\label{l3}\\
 n_adx^a&=&dr+Qd\zeta+\overline{Q}d\overline{\zeta}-H\ell_adx^a\label{n3}\\
m_adx^a&=&\frac{\sqrt{2}}{\rho P}d\overline{\zeta}. \label{m3} \eea
Once we have the radial dependence of $Q$ and $H$, we have the
radial dependence of the metric. From the $[\Delta, D]$ commutator
we find
$$DH=-(\gamma+\overline{\gamma}),$$
while from (A.3f) and (A.4c)
\begin{eqnarray}
D\gamma&=&\psi_2,\label{g1}\\
D\psi_2&=&3\rho\psi_2,\label{g2}
\end{eqnarray}
so that, by (\ref{r1}),
\begin{eqnarray}
\psi_2&=&\rho^3\psi_2^0,\label{g3}\\
\gamma&=&\gamma^0+\frac{1}{2}\rho^2\psi_2^0,\label{g4}
\end{eqnarray}
where $\psi_2^0$ and $\gamma^0$ are independent of $r$. Therefore
\beq \label{h1}H=H^0-(\gamma^0+\overline{\gamma}^0)r
-\frac{1}{2}\rho\psi_2^0-\frac{1}{2}\overline{\rho}\overline{\psi}_2^0,
\eeq
where $H^0$ is independent of $r$; so (\ref{h1}) gives the radial
dependence of $H$.

For $Q$, the commutator $[\delta, D]$ gives
$$-\frac{\overline{\rho}P}{\sqrt{2}}DQ=\overline{\alpha}+\beta,$$
while (A.3d) and (A.3e) can be integrated to give
$$\alpha=-\alpha^0\rho\;\;;\quad \beta=-\beta^0\overline{\rho},$$
with $\alpha^0$ and $\beta^0$ independent of $r$. Therefore
\beq\label{q1}
Q=Q^0+\frac{\sqrt{2}}{P}(\overline{\alpha^0}+\beta^0)r, \eeq
with $Q^0$ independent of $r$.

For later use, we find the radial dependence of the remaining spin
coefficients, $\mu$ and $\nu$. For $\mu$, we integrate (A.3h) to
find
\beq
\mu=\mu^0\overline{\rho}+\frac{1}{2}\rho(\overline{\rho}+\rho)\psi^0_2,
\label{h2}
\eeq
where $\mu^0$ independent of $r$. For $\nu$, first from (A.4e), assuming $\Phi_{12}=0$,
\[\psi_3=\psi^0_3\rho^2+\psi^1_3\rho^3+\psi^2_3\rho^4,\]
where $\psi^i_3$ are independent of $r$, and then from (A.3i),
\beq
\nu=\nu^0+ \psi^0_3\rho+\frac{1}{2}\psi^1_3\rho^2+\frac{1}{3}\psi^2_3\rho^3,
\label{h3}
\eeq
where $\nu^0$ is independent of $r$.

\medskip

We now note the changes in the  non-vacuum case. As in the
Goldberg-Sachs theorem, we continue to insist on
\[\Phi_{00}=\Phi_{01}=\Phi_{02}=0=\Lambda,\]
but allow the possibility of  non-zero $\Phi_{11}$,
$\Phi_{12}$, and $\Phi_{22}$  (in fact, $\Phi_{22}$ doesn't
arise in the calculation). This changes some of the details
above. We still have
\[\kappa=\sigma=\epsilon=\tau=\pi=\lambda=0,\]
and (\ref{r1}), but (\ref{g1})-(\ref{g4}) change.  For the radial
dependence of $\Phi_{11}$ we have equation (A.4i):
\[D\Phi_{11}=2(\rho+\overline\rho)\Phi_{11},\]
which integrates readily.
With this, (A.4c) can be integrated for $\psi_2$, then (A.3f) for
$\gamma$ and then $H$ obtained from the commutator $[\Delta,D]$. In
place of (\ref{g3})-(\ref{g4}) and (\ref{h1}), we find
\bea
\Phi_{11}&=&(\rho\overline\rho)^2\Phi^0_{11}\label{ng1}\\
\psi_2&=&\rho^3\psi^0_2+2\rho^3\overline\rho\Phi^0_{11}\label{ng2}\\
\gamma&=&\gamma^0+\frac{1}{2}\rho^2\psi_2^0+\rho^2\overline\rho\Phi^0_{11}\label{ng3}\\
H&=&H^0-(\gamma^0+\overline{\gamma}^0)r
-\frac{1}{2}\rho\psi_2^0-\frac{1}{2}\overline{\rho}\overline{\psi}_2^0-\rho\overline\rho\Phi^0_{11}.\label{ng4}
\eea
where, as usual, quantities with a superscript  zero are independent
of $r$, and $\Phi^0_{11}$ is real. This is enough for the metric.
For the spin-coefficient $\mu$, (A.3h) now gives
\be
\mu=\mu^0\overline{\rho}+\frac{1}{2}\rho(\overline{\rho}+\rho)\psi^0_2+\rho^2\overline\rho\Phi^0_{11}.
\label{ng5}
\ee
The remaining spin-coefficient $\nu$ is altogether more complicated.
We need to solve (A.4j) for $\Phi_{12}$, then (A.4e) for $\psi_3$
and then (A.3i) for $\nu$. The results are polynomials in $\rho$ and
$\overline\rho$, with coefficients constant along $\ell$. We don't
need the detailed expressions for these quantities, which can be
found in \cite{TrimWainwright}.  For our purposes, the following
suffices
\bea
\Phi_{21}&=&O(|\rho|^3)\;, \nonumber\\
\psi_3&=&O(|\rho|^2)\;, \nonumber\\
\nu&=&\nu^0+O(|\rho|)\label{ng6}\;. \eea

\medskip

We are ready to prove now:

\begin{Proposition}
\label{Pcoords} Let
\beaa
  \mcN  &=&\{p\in \mcM: \ \mbox{ the null geodesic through $p$}
  \\
  && \ \mbox{
tangent to $\ell(p)$ has an end point on $\Spti$}\}
 \;.
\eeaa
There exist coordinates $(u,r,\zeta)$ parameterizing a neighborhood
of $\mcN $, such that $(u,\zeta)$ coincide with Bondi coordinates on
$\Spti$, in which the metric takes the form $2\ell_{(a}n_{b)}- 2
m_{(a}\bar m_{b)}$, with $\ell$, $n$ and $m$ given by
\eq{m1}-\eq{m3}, with $\rho$ given by \eq{r1}, $H$ given by \eq{h1},
while $Q$ is given by \eq{q1}, where $\alpha^0$, $\beta^0$,
$\psi^0_2$, $H^0$, $Q^0$, $L$ and $\Sigma$ are smooth functions of
$u$ and $\zeta$.
\end{Proposition}

\proof  Transversality and smoothness of $\tell$ at $\Spti$ imply
that there exists a neighborhood of $\Spti$ on which  $\tell$ is
transverse to $\scrip$, and the result follows from smoothness of
$\ell$ together with the calculations above. \qed

 Now we have the $r$-dependence of
the metric. By construction, the coordinates $u$ and $\zeta$ are
good coordinates on $\scripo$  near $S^+$, while $r\Omega\rightarrow
1$. Rescaling the metric by $r^{-2}$ and setting $R=r^{-1}$, we
obtain for the asymptotic behavior
\bel{scripcomp}
 \tg=R^2g=2(du-\sqrt{2}\frac{L}{P}d\zeta
 -\sqrt{2}\frac{\overline{L}}{P}d\overline{\zeta})(-dR+O(R))-4\frac{d\zeta
 d\overline{\zeta}}{P^2}(1+O(R^2))
 %\;,
\ee
(recall that we shifted $r$ to obtain $r^0=0$), which shows
explicitly that the space-time is weakly asymptotically simple with
this choice of rescaling.

Let
$
 \Smti $
%
%denote the image of the intersection of $\overline {\mcN}_i$
%with $\scrim$ under this embedding, thus $\Smti$ is
be obtained by flowing $\Spti$ from $\scrip$ to $\scrim$ along
$\tell$. We have:

\begin{Lemma}
\label{LBILY} $\Smti$ is a smooth acausal cross-section of $\scrim$,
with both $S^+$ and $S^-$  diffeomorphic to $S^2$.
\end{Lemma}

\proof In the construction leading to \eq{scripcomp} we consider
instead $r\to-\infty$, taking $R=-r^{-1}$ to obtain a conformal
completion $\overline{\mcU_i}$ at past infinity of the coordinate
patch, say $\mcU_i$, constructed above. Consider the map $\psi$
which to $p\in \Spti$ assigns the generator of $\scrim= \R\times
S^2$ which is met by the maximally extended null geodesic tangent to
$\tell$ and passing through $p$.
Applying~\cite[Theorem~3.1]{ChConformalBoundary} to
$\overline{\mcU_i}$ we conclude that there exists a smooth local
diffeomorphism from $\overline{\mcU_i}$ to $\tmcM$, so that $\Smti$
is a smooth immersed submanifold of $\scrim$; note, however, that
$\Smti$ might fail to be embedded because some points of $\scrim$
could be met by more than one integral curve of $\tell$ emanating
from $\Spti$. In any case, we infer that $\psi$ is a local
diffeomorphism. By~\cite[Exercise 11-9, p. 253]{Lee:TM} $\psi$ is a
covering map, and since $S^2$ is simply connected it follows that
$\psi$ is a diffeomorphism, so $S^+\approx S^2$, and $\Smti$
intersects every generator precisely once. As the only  causal
curves within $\scrim$  are the generators of $\scrim$, the result
follows.
\qed

As observed by Mason~\cite{Mason:AlgSpec}, one has

\begin{Proposition}
\label{Pmasschange}
 The Trautman-Bondi mass $\mTBp$ of $\Spti$ equals the negative of
 the Trautman-Bondi mass $\mTBm$ of $\Smti$.
\end{Proposition}

This will follow if the mass aspect has the same property.  Mason
suggests two proofs for this proposition: either via a direct check
on the mass aspect or by exploiting an alternative formula for the
Trautman-Bondi mass given in \cite{PenroseRindlerv12}. We shall
present the first, exploiting a formula in \cite{NatorfTafel} for
the mass aspect. First we note that the correspondence
$(u,r,\xi)_{NT}=(u,r,\zeta\sqrt{2})_{CT}$ relates our coordinates
(subscript CT) to the ones used in \cite{NatorfTafel} (subscript
NT). Then the quantities $(L,H,W,m+iM,\Sigma,\hat{P})_{NT}$ arising
in their metric are for us $(-\frac{L}{P}, -H, \frac{Q}{\sqrt{2}},
\psi_2^0,\Sigma,1)_{CT}$, and finally their operator $\partial$
translates for us as
\be %l{Tafelform}
 \partial=\frac{1}{\sqrt{2}}\left(\frac{\partial}{\partial\zeta}+\frac{L\sqrt{2}}{P}\frac{\partial}{\partial
 u}\right).
\ee
%\]
%
With these preliminaries,  the  Natorf-Tafel formula (47) of
\cite{NatorfTafel} for the integrand of the Trautman-Bondi mass
(which we will refer to as the \emph{mass aspect}; but note that
this is \emph{not} the original mass aspect function
of~\cite{BBM,Sachs})
 translates for us to
\bel{Tafelform}
{\mathcal{M}}=\frac{1}{2}(\psi_2^0+\overline\psi_2^0)+3\Sigma\Sigma_{,u}
+\frac{1}{2}(\tilde\Delta+2)\eta-2iP(\overline{L}_{,u}\partial\Sigma-L_{,u}\overline{\partial}\Sigma)
 \;, \label{mas}
\eeq
where $\tilde\Delta=P^2(\partial\overline\partial+\overline\partial\partial)$ and
\[\eta=-\frac{1}{2}P^2\left(\partial\left(\frac{\overline{L}}{P}\right)+\overline{\partial}\left(\frac{L}{P}\right)\right).\]
To investigate the mass  aspect at $\scri^-$, we define a
time-reversed space-time for which the old $\scri^-$ is now
$\scri^+$. With hatted quantities referring to the time-reversed
space-time, we take coordinates $(\hat u, \hat r,
\hat\zeta)=(-u,-r,\overline\zeta)$.
(The redefinition $r\to -r$ should be clear in our context;  the
need to replace $\zeta$ by $\overline \zeta$ arises then from
elementary orientation considerations; the transition $u\to -u$
arises from the fact that we will be using on $\scrim$ a formula for
the Trautman-Bondi mass which has been worked out at $\scrip$, and
this requires a change of time-orientation). Then all our
calculations so far can be repeated in the tetrad $(\hat\ell, \hat
n, \hat m)=(-\ell,-n,\overline m)$.  In particular $\hat\rho$ equals
$-\overline\rho$, and tracing through the quantities in the mass
aspect, we find $(\hat\Sigma,\hat L, \hat\eta,\hat\partial,
\hat{\psi}_2^0)=(\Sigma,-\overline L,-\eta,\overline\partial,
-\overline{\psi}_2^0)$. Using these in (\ref{mas}), we see that, as
desired, ${\mathcal{M}}$ changes sign. Now all quantities appearing
in (\ref{mas}) are constant along $\ell$, and we conclude that the
mass aspect at $\Smti$ is the negative of the mass aspect at
$\Spti$.
\qed

\section{Spacelike hypersurfaces, a rigid positive energy theorem}
\label{Scss}

 Choose a cut $\Spto $ of $\scrip$
and consider the associated null boundary
\bel{defN}
 \mcnt :=\dot J^-(\Spto ,\mcM)
 \;,
\ee
%,
then %, by Lemma~\ref{LBILY},
$\mcnt $ is an achronal hypersurface generated by null geodesics
orthogonal to $\Spto $. Further there exists a neighborhood $\mcO$
of $\scrip$ such that $\mcnt \cap \mcO$ is smooth. If we assume
$(\mcM,\fourg)$ to be  globally hyperbolic, there exists a
time-function $\tau$ on $\mcM$ with the property that its level
sets,
$$
 \zhyp_{\tau_0}:=\{\tau=\tau_0\}\;,
$$
are smooth spacelike Cauchy surfaces (compare~\cite{BernalSanchez}).
Define
$$
 \mcnt _\tau:= \dot J^-(\Spto ) \cap \zhyp_\tau
 \;;
$$
note that the intersection is transverse. Since $\mcM= \cup_\tau
\zhyp_\tau$, we have that $\cup_\tau \mcnt _\tau=\mcnt $. This,
together with Dini's theorem, shows that there exists $\tau_0$ such
that $\mcnt _{\tau_0} \subset \mcO$, thus $\mcnt _{\tau_0}$ is a
smooth sphere.

For $\epsilon>0$ let $\tgf_\epsilon$ be a family of smooth
Lorentzian metrics on $\tmcM $ such that $\tgf_\epsilon$ converges
to $\tfourg$ on compact subsets of $\tmcM $ as $\epsilon$ goes to
zero in the $C^\infty$ topology, with the property that all vectors
which are null for $\tgfe$ are spacelike for $\tfourg$. By
continuous dependence of geodesics upon the metric, for $\epsilon>0$
small enough  all null $\tgfe$-geodesics normal to $\scrip$
intersect $\zhyp_{\tau_0}$ in a smooth sphere $\mathring{N}$, with
the corresponding hypersurface $\mcnt ^\epsilon$, defined as in
\eq{defN} using the metric $\tgfe$, being smooth in its portion
which is bounded by $\Sp$ and by $\mathring N$; call this region
$\hyp_\ext$. The Cauchy surface $ \zhyp_{\tau_0}$ is contractible by
one of the hypotheses of Theorem~\ref{Tm1v0}, or
by~\cite[Section~5]{NewmanAF} if full asymptotic simplicity is
assumed. Simple connectedness of $\zhyp_{\tau_0}$ and elementary
intersection theory show that $\mathring N$ separates
$\zhyp_{\tau_0}$ into two components. From the Hurewicz isomorphism
theorem~\cite[Chapter~7, Section~5]{Spanier} we further conclude
that $H_2(\zhyp_{\tau_0})$ is trivial, which implies  that one of
the components separated by the sphere $\mathring N$, say $\mcK$, is
compact. Set
$$
 \hypo=\mcK\cup \hyp_\ext
 \;,
$$
% be the
then $\hypo$ is a piecewise differentiable $\fourg$--spacelike
hypersurface which is the union of a compact set and of an
asymptotic region extending to $\scrip$. Smoothing out the corner at
$\mathring N$ one obtains a smooth hypersurface, still denoted by
$\hypo$. Next, the formulae of~\cite[Appendix~C.3]{CJK}  show how to
make a small deformation of $\hypo$ near $\scrip$ to obtain a
hypersurface on which the induced metric asymptotes to a hyperbolic
one,
as needed for the proof of positivity of mass of~\cite{CJL}.
Finally, we let $ \hhypo$ be the universal cover of $\hypo$, then
$\hhypo$ is complete, with one or more asymptotically hyperbolic
ends.%
\footnote{Conceivably one can infer at this stage, from the
results in~\cite{NewmanAF}, that $\hypo$ is simply connected;
however, the argument that follows sidesteps this issue.}

By the positive energy theorem of~\cite{CJL} applied to a chosen
asymptotic end of $\hhypo$ we conclude that the Trautman-Bondi mass
associated with this end is non-negative. An identical construction
starting from $\Smti$ shows that $\mTBm\ge 0$. From
Proposition~\ref{Pmasschange} we infer that
$$
 \mTBp=0
 \;.
$$

We continue with an investigation of the consequences of the
Witten-type proof of the positive energy theorem on $\hhypo$. Let
$\zpsi$ be a Dirac spinor which is parallel with respect to the
spin-connection associated with the Minkowski metric, such that the
resulting Killing vector in Minkowski space-time $\R^{1,3}$ equals
$\partial_t$.
Now, when $m=0$, the proof of the positive energy theorem
in~\cite{CJL} shows that the space-time metric $\fourg$ is flat
along $\hhypo$, and that there exists a spinor $\psi$, solution of
the Witten equation, such that $\psi=\zpsi+\chi$, with $\chi$ in a
weighted Sobolev space obtained by completing $C_0^\infty$ with
respect to the norm
$$
 \sqrt{\int |D\chi|^2 d\mu_g} \;.
$$
Furthermore, $\psi$ is parallel with respect to the space-time
connection $\nabla$ associated to the initial data set $(\hhypo,
g,K)$,
\bel{consspi} \nabla_i\psi:=
 D_i \psi +\frac 12 K_{ij}\gamma^j\gamma^0 \psi = 0
 \;.
\ee

Let $(V,Y)$ be the KID defined by $\psi$,
$$
V:= \langle \psi, \psi\rangle \;, \quad Y:=\langle \psi,
 \gamma^0\gamma^j\psi\rangle e_j \;.
$$
\Eq{consspi} implies that $(V,Y)$ is parallel, in the following
sense:
\bel{consvec}
 D_i V= K_{ij}Y^j\;, \quad D_i Y_j = V K_{ij} \;.
\ee
It follows that the Lorentzian norm squared $V^2 - |Y|_g^2$ of
$(V,Y)$ is constant on $\hyp$,
\bel{consvec2} D_i\Big(V^2 - |Y|_g^2\Big) = 0
 \;.
\ee
It should follow from the methods in~\cite{AndChDiss} that this norm
is strictly positive by choice of $\zpsi$, so that the associated
Killing vector is timelike; however, an argument which avoids the
heavy machinery of the last reference proceeds as follows: If $V^2 -
|Y|_g^2 = 0$, we choose a different asymptotic value of $\zpsi$. If
the new resulting Killing vector is timelike we are done, otherwise
there is a linear combination of the new Killing vector  and of the
old one which is timelike, and satisfies \eq{consvec}, leading to a
timelike Killing vector for which \eq{consvec2} holds as well.

 We consider the Killing
development $(\zM,\zg)$ defined by $(\hhypo,V, Y)$, thus $ \zM$ is
$\R_t\times \hhypo$ with metric
$$
 \zg =  V^2 dt^2 - g_{ij}(dx^i+Y^i dt)(dx^j+Y^j dt)
 \;,
$$
with Killing vector $X=\partial_t$.
 Letting
\bel{mubound2} \exp({\mu}):=\;  ^4\!g(X,X) = V^2 - |Y|^2_g
 \;,
\ee
we rewrite the Killing development metric $\zg$ in the following
form,
\bel{gfourm} \zg =  \exp(\mu)(dt+ {\theta_idx^i} )^2-\threeg
 \;,
\ee
as needed in Lemma~3.11 of~\cite{CMT}, where the Riemannian metric
$\threeg$ is related to the initial data metric $g$ by the equation
\bel{hmet}
 \threeg_{ij}= g_{ij}+ \exp(\mu)\theta_i\theta_j
 \;,
\ee
%.
and
\bel{thetaform}
 \theta_i = -e^{-\mu}g_{ij}Y^j
 \;.
\ee
To apply that last Lemma, we need to verify that the metric $h$ in
\eq{gfourm} is complete, and that the $h$-length of $\theta$ is
uniformly bounded on $\hhypo $. Now, the hyperbolic asymptotics  of
$g$, together with compactness of $ \hhypo \cup \,\,\dot{\!\! \hhypo
}$ and the Hopf-Rinow theorem, imply completeness of $(\hhypo ,g)$.
Since the last term in \eq{hmet} gives a non-negative contribution
on any given vector, completeness of $(\hhypo ,\threeg)$ follows
from that of $(\hhypo ,g)$.

Next, it follows from \eq{consvec2} that $\exp(\mu)$ is constant
over $\hhypo $. Further,
$$
h^{ij} =g^{ij}- \frac{\exp{(-\mu)}}{1+\exp{(-\mu)}|Y|_g^2}Y^iY^j
 \;,
$$
so that
$$
|\theta|_h^2=h^{ij}\theta_i\theta_j =
\frac{\exp{(-2\mu)}|Y|_g^2}{1+\exp{(-\mu)}|Y|_g^2}
 \le \exp(-\mu)=:C
 \;,
$$
which establishes the desired uniform bound on $ |\theta|_\threeg$.

From~\cite[Lemma~3.11]{CMT} we conclude that
the Killing development $(\zM,\zg)$ of $(\hhypo,V,Y)$ is
geodesically complete. Since $\hhypo$ is simply connected, so is
${}\zM\approx \R\times \hhypo$. The Lorentzian version of the
Hadamard-Cartan theorem~\cite[Proposition~23, p.~227]{BONeill}
implies that $(\zM,\zg)$ is $\R^{1,3}$. In particular $\hhypo$ is a
hyperboloidal hypersurface in $\R^{1,3}$, and hence has only one
asymptotically hyperbolic end. But if $\hypo$ were not simply
connected, $\hhypo$ would have had more than one such end. We
conclude that $\hhypo=\hypo$.

By hypothesis $(\mcM,\fourg)$ satisfies the dominant energy
condition, hence  the domain of dependence $\mcD(\hypo,\mcM)$
in the original space-time $\mcM$ is vacuum
by~\cite[Section~4.3]{HE}. From~\cite{CBG} we conclude  that
$\mcD(\hypo,\mcM)$  is isometrically diffeomorphic to a
globally hyperbolic subset of the domain of dependence
$\mcD(\hypo,\zM)$ in the Killing development. This is a
bijection when $(\mcM,\fourg)$ is both past and future
asymptotically simple, otherwise $\mcD(\hypo,\mcM)$ couldn't be
future null geodesically complete.

For the record, the above establishes the following rigidity
statement (compare Theorems~5.4 and 5.7 of~\cite{CJL}; the reader is
referred to this last reference for precise definitions):

\begin{Theorem}
\label{Tpos} Let $\mu$,  respectively $J^i$, denote the energy
density, respectively the momentum density, of an initial data set
$(\hyp,g,K)$.   Suppose that $(\hyp,g)$ is geodesically complete
without boundary, and that $\hyp$ contains an end which is
$C^4\times C^3$, or $C^1$ and polyhomogeneously, compactifiable and
asymptotically CMC, with energy-momentum density decaying fast
enough.  If
\bel{dec}\sqrt{g_{ij}J^iJ^j} \le  \mu
 \;,
\ee
and if the Trautman-Bondi mass of $\hyp$ vanishes, then $(\hyp,g,K)$
can be realized by embedding $\hyp$ into Minkowski space-time.

If the initial data set is known to be vacuum near the conformal
boundary from the outset, or to satisfy a set of equations which are
well behaved under singular conformal transformations such as, e.g.,
the Einstein--Maxwell or Einstein--Yang-Mills equations,   then the
restriction that the data be asymptotically CMC is not needed.
\qed
\end{Theorem}

\begin{Remark}
\label{RTpos} It is still an open question whether a null
Trautman-Bondi energy-momentum is compatible with the remaining
hypotheses above; it would be of interest to settle that.
\end{Remark}

We continue by pushing $\Sp$ slightly down the generators of
$\scrip$, to conclude that the space-time metric is flat in a
neighborhood of $\Sp$. By~\cite{ChConformalBoundary} we conclude
that

\begin{Proposition}
 \label{Pfact}
There exists a neighborhood of $\Sp$ which is isometrically
diffeomorphic to a neighbourhood of a spherical cut of $\scrip$ in
Minkowski space-time. Moreover the space-time metric is flat to the
future of any spacelike hypersurface spanned by $\Sp$.
\qed
\end{Proposition}

Recalling that the congruence generated by $\ell$ has nowhere
vanishing twist (see Proposition~\ref{Pnotwist}),
Theorem~\ref{Tm1v0} follows now from Proposition~\ref{Pnogo}:
\qed

\begin{Proposition}
\label{Pnogo} There exists no smooth, null, geodesic congruence
defined in a neighborhood of a cross-section $\Sp$ of the
Minkowskian $\scri$ which is shear-free, transverse to $\Sp$, and
has nowhere vanishing twist.
\end{Proposition}

\proof
Any smooth, null-geodesic congruence near a Minkowskian $\scrip$
defines a spin-weight one function $L$ as in \eq{o1}
which is smooth on $\scrip$ in a neighbourhood of the cut
$\Sp=\{u=0\}$. By~\cite[Equation~(2.24)]{LindNewman} the shear-free
condition in Minkowski space is equivalent to
 \be\label{sf1} \eth
 L+L\dot{L}=0 \;,
  \ee
where the dot stands for $\partial/\partial u$ and
$\eth$ is the eth-operator of Newman and Penrose.
%, defined on a
%function of spin-weight $s$ by
%\[\eth F=P^{1-s}\frac{\partial(P^sF)}{\partial\zeta}.\]
Then $\Sigma$ in \eq{r1} is given by (see \cite[Equation
(2.17)]{LindNewman})
\be
 \label{sf2} \Sigma= \frac i 2 (\overline\eth
L+\overline{L}\dot{L}-\eth\overline L-L\dot{\overline{L}})
 \;.
\ee
Now consider $F:=L\overline{L}$ restricted to the cut $u=0$. Clearly
$F$ has a maximum on this sphere, and at the maximum its gradient
vanishes, so at any maximum
\be\label{sf3} 0=\eth F=(\eth L
)\overline{L}+L(\eth\overline{L})=L(\eth\overline{L}-\overline{L}\dot{L}),
\ee
using (\ref{sf1}) to eliminate $\eth L$. If $L\equiv 0$ on the cut
$\{u=0\}$ then $\Sigma=0$ throughout the cut by \eq{sf2}, and we are
done. Otherwise, at a maximum of $F$, $L$ does not vanish so that
the second factor in (\ref{sf3}) must. But by (\ref{sf2}) this
forces $\Sigma$ to vanish there.
\qed

% \ptc{dangerous renew commands for Spti and scripo etc }
\renewcommand{\hhypo}{\Hhypo}
\renewcommand{\hypo}{\hyp_i}
\renewcommand{\Smti}{\Sm_i}
\renewcommand{\Spti}{\Sp_i}
\renewcommand{\Spto}{\Sp_i}
\renewcommand{\scripo}{\scrip_0}

{\noindent\sc Proof of Theorem~\ref{Tnew}:}
Under \eq{newcond}, the argument of the proof of
Theorem~\ref{Tm1v0} with $\Sp$ replaced by $\Spti$ carries
through with minor modifications. Indeed, the cross-sections
$\Smti$ are smooth acausal cross-sections of $\scrim$
  as before, because they are constructed
by flowing along the geodesics which start at $\scrip$, and
those do not care about smoothness of $\ell$ as a vector field
on $\mcM$. Next, the argument that the Bondi mass aspect
changes sign remains valid for those members of the congruence
which have end points on $\oSpti$. But the Bondi mass aspect is
a smooth function both on $\Spti$ and $\Smti$, and the density
hypothesis \eq{newcond} guarantees that the corresponding
subset of $\Smti$ is dense. Continuity allows us to conclude,
as before, that the Trautman-Bondi mass changes sign when
replacing $\Spti$ with $\Smti$.

 Let $\hypo$ be a
hypersurface $\hyp$ as in the proof of Theorem~\ref{Tm1v0} with
$\Sp$ there replaced by $\Spti$. We have shown so far that $\hypo$
is a hyperboloidal hypersurface in Minkowski space-time and, since
it has no edge, its future (whether in Minkowski space-time or in
$\mcM$) coincides with its future domain of dependence there:
\bel{ddeq}  \mcD^+(\hypo ,\mcM)= J^+(\hypo ,\mcM)
 \;.
\ee
Now, by asymptotic simplicity, every generator of the Cauchy
horizon%
\footnote{There are two conventions for defining
$\mcD(\hyp)$, we use the one in which inextendible timelike curves
are required to intersect $\hyp$ precisely once.}
$\dD ^-(\hypo)$ has a future end point on $\Spto$.
This implies that
\bel{boundeq}
 \dD ^-(\hypo,\mcM )=\dot J^-(\Spti,\tmcM )\cap \mcM
 \;.
\ee
%
%so that
%%
%$$
% \mcD^-(\hypo,\tmcM )= \overline{J^-(\hypo,\tmcM )\setminus
% J^-(\Spti,\tmcM )}
% \;.
%$$
%% .
%We conclude that for every $i\in\N$
% of general fact anyway?}
%%
%\bel{flatd}
% \mcD(\hypo,\mcM )=J^+(\dot J^-(\Spti,\tmcM )\cap \mcM)
% \;.
%\ee
%%
%and that the space-time metric $g$ is flat in this set.

We continue by showing that%
%%
%$$
% \underbrace{\cup_{t\in (-\infty,\tau_0]}\mcD(\hyp_t)}_{=:\myset} = \mcM
% \;.
%$$
%%
%
\bel{equality}
 \underbrace{\cup_{i\in\N}\mcD(\hypo,\mcM)}_{=:\myset} = \mcM
 \;.
\ee
Suppose that this is not the case, then there exists a sequence of
points $p_i \in \dot\mcD^-(\hypo)$ which converges to a point $p$
belonging to the boundary $\dU $ of $\mcU$. Let $\dot\gamma_i$ be
the vector tangent to a generator of $\dD ^-(\hypo)$ at $p_i$,
normalized to unit norm with respect to an auxiliary Riemannian
metric. Passing to a subsequence if necessary, the sequence $(\dot
\gamma_i)$ converges to a null vector $\dot \gamma$ at $p$. Let
$\gamma$ be a null geodesic through $p$ with tangent $\dot \gamma$
there, maximally extended in $\tmcM $, then $\gamma$ meets $\scrip$
at some point $q$.

Without loss of generality, passing to a subsequence if necessary,
we can assume that
$$
  \Sptimo\subset J^+(\Spti,\tMM)
  \;.
$$
Since $\scrip\subset \cup_i J^+(\Spti)$ there exists $\io$ such that
$q\in J^+(\Sptio)$. Then $\gamma$ intersects $\mcD(\hypi)$ for every
$i<\io$, and since $p\not \in \mcD(\hypi)$ the null geodesic
$\gamma$, when followed to the past starting from $q$, has to
intersect $\dot J^-(\Spti)$ before reaching $p$, compare
\eq{boundeq}. But the $\gamma_i$'s accumulate at $\gamma$ as $i$
tends to infinity, so that there exists $i_1>i_0+1$ so that (by
continuous dependence of solutions of ODE's upon initial values) the
geodesic $\gamma_{i_1}\subset \dot J^-(S^+_{i_1})$ intersects $\dot
J^-(S^+_{i_0+1})$. This is, however, not possible since $\dot
J^-(S^+_{i_0+1})$ is strictly interior to $J^+(\dot
J^-(S^+_{i_1}))$. We conclude that \eq{equality} holds, and
therefore $\fourg$ is flat.

Summarising, $(\mcM,\fourg)$ is a simply connected, flat,  null
geodesically complete manifold. Theorem~\ref{Tnew} follows now from
Proposition~\ref{Pflatr} below.%
\footnote{We are grateful to a referee for a
suggestion leading to Proposition~\ref{Pflatr}.}
\qed

\begin{Proposition}
 \label{Pflatr}
 Let  $n\ge 1$.  The only $(n+1)$-dimensional simply connected, flat,
null \underline{or} timelike geodesically complete Lorentzian
manifold $(\mcM,g)$ is, up to isometric diffeomorphism, the
Minkowski space-time $\R^{1,n}$.
\end{Proposition}

\proof
Since $ g$ is flat, the dimension of the set of germs of locally
defined Killing vector fields is the same at every point. A theorem
of Nomizu~\cite{Nomizu} shows then that every local Killing vector
extends to a globally defined one.
By~\cite[Lemma~1]{GarfinkleHarris}, all  Killing vector fields%
\footnote{The hypothesis that the Killing vector is timelike, made
elsewhere in~\cite{GarfinkleHarris}, is not used in the proof, which
goes through unchanged  with one exception: when $n=1$, and the
manifold is assumed to be null geodesically complete, and  the
Killing orbit is null. But in this case the  orbit is a null
geodesic, so null geodesic completeness implies completeness of that
orbit trivially.}
are complete. But, in a flat space-time, affinely parameterized
geodesics are orbits of translational Killing vectors, hence
$(\mcM,\fourg)$ is geodesically complete. The result follows
now from the Hadamard-Cartan theorem.

\qed

\section{Concluding remarks}
\label{Sconcl}

One would like to remove all restrictive hypotheses of
Theorem~\ref{Tnew} and assert that the only algebraically special
vacuum asymptotically simple space-time is the Minkowski one. Any
proof of this,  in a setting where the set
\beaa
 \mcV&:=&\{p\in \Scrip | \
 \mbox{$p$ is an end-point of an integral curve $\gamma$ of
 $\ell$}\}\subset \Scrip
\eeaa
does \emph{not} cover a dense subset of some sequence of
cross-sections of $\scrip$, has to use arguments going  beyond those
indicated  by Mason. On the other hand, one could expect that some
version of the current argument should apply if the last density
property holds. However, attempts to include such situations face
several difficulties. Suppose, for instance, that $S$ is a
cross-section of $\scrip$ such that $\mcV\cap S$ is dense in $S$.
Now, Mason's construction requires flowing from $ \mcV\cap S$ to the
past along the integral curves of $\tell$. Since $ \mcV\cap S$ is
not compact anymore, the geometry of the resulting subset $S^-$ of
$\scrim$ is not clear: By causality considerations, $S^-$ will be
bounded to the future on $\scrim$, however, it could very well be
unbounded to the past. Regardless of that issue, the closure
$\overline {S^-}$ of $S^-$ might fail to be differentiable. Finally,
$S^-$ might develop self-intersections. In all those cases a useful
notion of   mass of $S^-$ is not clear, and certainly no suitable
positivity theorem is available. Similar problems concerning the
geometry of $\Smti$ could occur in those space-times in which
$\scrim$ is not diffeomorphic to $\R\times S^2$; while we are not
aware of any such asymptotically simple examples, their existence
has not been ruled out so far (strong causality at $\scrim$ must
then necessarily fail, compare~\cite{NewmanAF}).

In this context the following example is rather instructive:
Consider a cut $S$ of $\scrip$ in Minkowski or  Schwarzschild
space-time given by the equation $u=\alpha$, then the integrand of
the Trautman-Bondi mass   of $S$ equals
$$
 \frac m {4\pi}+ \frac 1 {16\pi}\Delta_2(\Delta_2+2) \alpha
 \;,
$$
where $m$ is the Schwarzschild mass parameter (which we set to zero
in the Minkowski case), while $\Delta_2$ is the Laplacian on $S_2$
(see, e.g.,~\cite[p.~136]{CJK}).  Now, with a little work one finds
that for any $c\in \R$ the function
$$
\alpha_c=\frac
 c 4 \Big( \cos\theta\ln\tan(\theta/2)-2\ln\sin\theta\Big)
$$
is a solution of
\bel{keq}
 \Delta_2(\Delta_2+2)\alpha_c=c
\ee
%$$
%
away from the north and south poles. We can add to $\alpha_c$
elements of the kernel of the operator appearing in \eq{keq}
which, when allowing functions which are singular at the poles,
contains $ \ln\tan(\theta/2)$. By adding to $\alpha_c$ an
appropriate multiple of this last function one can obtain a
function $\alpha_{c,S}$ which solves \eq{keq} away from the
south pole, as well as a function $\alpha_{c,N}$ which is a
solution away from the north one
$$
 \alpha_{c,N}= \alpha_c - \frac {c} 4 \ln \tan(\theta/2)
 \;,\qquad
 \alpha_{c,S}= \alpha_c + \frac {c} 4 \ln \tan(\theta/2)
 \;.
$$
The graph $\{u=\alpha\}$ of each of these functions provides thus an
example of an embedded smooth submanifold of $\scrip$ (which fails
to be a cut of $\scrip$ because it misses one generator) with
Trautman-Bondi mass $m_{TB}$, when naively defined as the integral
of the mass aspect function, being an affine function of $c$, in
particular both the mass aspect and $m_{TB}$ can be negative.

A piecewise smooth, non-differentiable, but (uniformly) Lipschitz
continuous cross-section of the Minkowskian $\scrip$, with mass
aspect function which is everywhere negative except at the equator
where it is not defined, can be constructed by using the function
$$
\alpha = \max(\alpha_{-1,S},\alpha_{-1,N})
 \;.
$$
The reader may readily devise a similar example in Schwarzschild
space-time, or in any space-time with a complete $\scrip$ in which
the relevant functions are uniformly bounded over $\scrip$.

\bigskip

\noindent{\sc Acknowledgements} We acknowledge useful discussions
with, or comments from,  M.~Anderson, G.~Galloway, W.~Natorf,
E.T.~Newman, J.~Tafel, and A. Trautman.

\appendix

\section{Smoothness of $\ell$ for non-branching metrics}
\label{Asmo}

Let $\ell$ be the field of principal null directions of the Weyl
tensor, normalized so that $\nabla_\ell \ell =0$. In this appendix
we wish to prove that $\ell$ is smooth on the set where the Weyl
tensor is non-branching, as defined in  the {introduction}; thus
either of type $II$ or $D$ throughout the set, or type $III$
throughout the set, or type $N$ throughout. As already mentioned, in
the type $II$ or $D$ case we allow the type to change from point to
point, as long as the Weyl tensor remains in the $II$ or $D$ class.

Since the claim is local, it is sufficient to establish the
result in a neighborhood of a point. So let  $o^A$, $\iota^A$
be any local basis of the space of two-component spinors near
$p$, and let $\psi_{ABCD}$ be the Weyl spinor. Then, \emph{by
definition}, the Weyl tensor is type $II$ or $D$ if at least
one of the solutions of the equation
\bean
 0&=& P(\lambda) :=\psi_{ABCD}(\lambda\iota^ A+ \omicron^A )(\lambda\iota^B
+ \omicron^B )(\lambda\iota^C +  \omicron^C )(\lambda\iota^D +
\omicron^D )
 \\
 & \equiv &
  \psi_ 4
\lambda^4 +\psi_3  \lambda^3 +\psi_ 2 \lambda^ 2+\psi_ 1\lambda
+\psi_0
 \eeal{psieq}
corresponds to a zero which is exactly of second order. The
associated principal null direction is  (whatever the type)
determined by the null vector $(\lambda\iota_A+
\omicron_A)(\overline\lambda\overline\iota_{ A'}+
\overline\omicron_{ A'})$. So smoothness of $\ell$ near $p$, for a
smooth metric, will be proved if we show that the solution $\lambda$
of \eq{psieq} depends smoothly upon the coefficients  $\psi_i$
appearing in \eq{psieq}. We will actually show that $\lambda$ is an
analytic function of the coefficients, see Proposition~\ref{PeigenW}
below, so $\ell$ will be analytic if the Weyl tensor is. The
analysis applies regardless of the order of the remaining roots of
\eq{psieq}, which explains why the argument covers both the $II$ and
$D$ Petrov-types (recall that type $II$ is defined by requiring the
remaining zeros to be simple, while type $D$ correspond to a second
order zero for the other root).

Similarly we define the Weyl tensor to be of type $III$ throughout a
set $\mcU$ if one of the solutions of \eq{psieq} corresponds to a
zero of exactly third order throughout $\mcU$; smoothness of the
associated vector field $\ell$ follows then from
Proposition~\ref{PeigenWk} below with $k=3$. Finally type $N$ is
defined by requiring $P$ to have one single zero of order four, and
smoothness is a consequence of Proposition~\ref{PeigenWk}  with
$k=4$.

We start by noting that,  by passing to a different basis of the
space of spinors if necessary, we can assume $\psi_4$ is non-zero at
$p$. Indeed, suppose that $\psi_4$ is zero in any basis at $p$, then
also $\psi_0=0$ for any basis at $p$, which implies $P(\lambda)=0$
for all $\lambda$. It follows that $\psi_i=0$ for all $i\in
\{0,\ldots,4\}$, hence $\psi_{ABCD}=0$ at $p$, thus the Weyl tensor
is of type $0$ there, contradicting our hypothesis that the Weyl
tensor is non-branching on the set under consideration. From now on
we choose any basis so that $\psi_4(p)\ne 0$, but then by continuity
there exists a neighborhood $\mcV_p$ of $p$ on which $\psi_4$ has no
zeros. All remaining considerations are restricted to $\mcV_p$,
which involves no loss of generality since $p$ is arbitrary within
the non-branching set.

Dividing by $\psi_4$, we are led to study the equation
\bel{poldef0}
 0 = \lambda^N + \sum_{i=0}^{N-1}\alpha_i \lambda^i\equiv W(\lambda)
 \;,
\ee
with smooth complex coefficients $\alpha_i$ (in the case of
current interest, $\alpha_i= \psi_{i}/\psi_4$, and $N=4$). Then
$\lambda$ is a zero of order two if and only if
$$
 W(\lambda)=W'(\lambda)=0\;, \quad \mbox{ but } \ W''(\lambda)\ne 0
 \;.
$$
We need to analyse the dependence of $\lambda$ upon the coefficients
$\alpha_i$ of \eq{poldef}. Consider, first, the equation
\bel{Weq} W'(\lambda)=0\;; \ee
the holomorphic implicit function theorem shows that \eq{Weq}
defines an analytic function $\lambda\equiv\lambda(\alpha_i)$ on the
set
\bel{Umcdef}
 \mcU_2:=\{W''(\lambda) \ne 0\;,\ \lambda \in\C\;,\ (\alpha_i)\in \C^N\}
 \subset \C^{N+1}
  \;,
\ee
%  $,
with
\bel{dereq}
 \frac{\partial\lambda}{\partial \alpha_i} = -\frac
 {i\lambda^{i-1}}{W''(\lambda)}
 \;.
\ee
Next, let the function $\varphi:\mcU_2\to \C$ be defined as
$\varphi=W(\lambda(\alpha_i))$, by definition we have
$W'(\lambda(\alpha_i))=0$ so that
\bel{calcy}
 d\varphi=   \frac{\partial\varphi}{\partial \alpha_i}
d\alpha_i = \Big( \underbrace{W'(\lambda)}_{=0}
\frac{\partial\lambda}{\partial \alpha_i} + \lambda^i\Big) d\alpha_i
=d\alpha_0 + \lambda d\alpha_1+\ldots+
 \lambda^{N-1}d\alpha_{N-1}
 \;.
\ee
It follows that $d \varphi$ has no zeros on $\mcU_2$, hence
$\{W(\lambda)=0\}$ is an analytic submanifold of $\mcU_2$.

We have thus shown
\begin{Proposition}
 \label{PeigenW} The set
$$
 \mcV_2:=\{\alpha_i:\ W(\lambda)=W'(\lambda)=0\;,\ W''(\lambda)\ne 0\ \mbox{ for some } \
 \lambda \in \C\}\subset \C^N
% \;.
$$
is an analytic submanifold of co-dimension one in $\C^N$, with
$\lambda$ being an analytic function on $
 \mcV_2$.
\end{Proposition}

The above generalizes immediately to zeros of $W$ which are
\emph{exactly of order $k$}: indeed, set
\bean
 \mcV_k
  &:= &
  \Big\{\alpha_i:\ \exists \ \lambda \in \C\ \mbox{such that} \ W^{(i)}(\lambda) =0\;,\ i=0,\ldots, k-1\;,
  \\
  &&
   \mbox{ but} \ \ W^{(k)}(\lambda)\ne 0 \Big\}\subset \C^N
 \;.
\eeal{Vdefk}
Then the equation $W^{(k-1)}(\lambda)=0$ defines a smooth function
$\lambda$ on $\mcV_k$ by the implicit function theorem, using an
obvious generalization of \eq{dereq}, and  for $k=1$ we are done.
Otherwise consider the map $\phi=(\phi^i):\mcV_k\to \R^k$, where
$$
 \phi^i=W^{(i)}(\lambda) \;,\quad i=0,\ldots,k-1
 \;.
$$
On the preimage $\phi^{-1}(\{0\})$ we have, as in \eq{calcy},
$\partial\phi^j/\partial \alpha_i = i(i-1)\cdots(i-j)\lambda^{i-j}$,
so that the last $k$ columns of the Jacobi matrix take the form%
%%
%$$
%\left(
%\begin{array}{cccc}
%  (k-1)\lambda^{k-2} &(k-2)\lambda^{k-3}
%                  & \cdots & 1 \\
%  (k-1)(k-2)\lambda^{k-3} & \cdots & 2 %i(i-1)\cdots(i-j)\lambda^{i-j}
%                                        & 0 \\
%  \vdots& \rotatebox{-15}{$\iddots$} & 0
%                  & 0 \\
%  (k-1)! & 0 & 0 & 0
%\end{array}
%\right)
% \;,
% $$
%%
%
$$
\left(
\begin{array}{cccc}
   \lambda^{k-1} &\lambda^{k-2}
                  & \cdots & 0! \\
   (k-1)\lambda^{k-2} & \cdots & 1! %i(i-1)\cdots(i-j)\lambda^{i-j}
                                        & 0 \\
  \vdots& %\rotatebox{-15}{$\iddots$}
  \iddots & 0
                  & 0 \\
  (k-1)! & 0 & 0 & 0
\end{array}
\right)
 \;,
 $$
the determinant of which is clearly non-vanishing. By the rank
theorem one concludes that:

\begin{Proposition}
 \label{PeigenWk} The set $\mcV_k$
is an analytic submanifold of co-dimension $k-1$
in $\C^N$, with $\lambda$ being an analytic function on $
 \mcV_k$.
\end{Proposition}

\begin{Remark}
Identical arguments apply to polynomials with real coefficients,
$\C$ being replaced by $\R$ and  ``analytic" being replaced by
``real analytic" both in the statements and in the proofs.
\end{Remark}

The argument just given also settles the following closely related
question:  consider a smooth function  $A:\mcU\to \End(\C^N)$ or
$A:\mcU\to \End(\R^N)$, defined on an open subset $\mcU$ of $\R^n$,
with the property  that for all $p\in \mcU$ %the map $A$ has  an
%eigenvalue with algebraic multiplicity $k$, with furthermore
the dimension of the associated eigenspace  equals $k$. We further
assume that $A$ is hermitian in the complex case, or symmetric in
the real one.
We claim that the function which to $p\in\mcU$ assigns the
associated $k$-dimensional eigenspace is a
smooth function on $\mcU$.%
\footnote{The question of multiple principal directions of the Weyl
tensor, discussed at the beginning of this section, can  also be
formulated as such a problem~\cite{Exactsolutions2}.}
In order to see this, let $\lambda$ be a solution of the
characteristic equation,
\bel{poldef}
 0 = \lambda^N + \sum_{i=1}^{N-1}\alpha_i \lambda^i\equiv W(\lambda):= \det(A-\lambda \Id)
 \;.
\ee
Then $\lambda$ will have algebraic multiplicity $k$ if and only if
the $\alpha_i$'s belong to the set $\mcV_k$ of
Proposition~\ref{PeigenW}.  Composing with the map which to $A$
assigns its symmetric polynomials $\alpha_i$, and using
Proposition~\ref{PeigenWk}, we conclude that $\lambda$ is a smooth
function on $\mcU$ (analytic if $A$ is).
 This allows us to show that:

\begin{Proposition}
\label{Peigsp}  The $k$-dimensional eigenspaces are smooth functions
on $\mcU$, analytic if $A$ is.
\end{Proposition}

\proof Let $p_0\in \mcU$ and let   $A_0=A(p_0)$,
$\lambda_0=\lambda(p_0)$, thus there exist $k$ linearly independent
vectors $e_i\in \C^N$ such that
$$
 (A_0-\lambda _0\Id)e_1=\cdots=
 (A_0-\lambda _0\Id)e_k=0
 \;.
$$
We can complete $\{e_i\}_{i=1}^k$ to  a basis $\{e_i\}_{i=1}^N$ of
$\C^N$. % so that $e_1=X_0$ and $e_2=Y_0$.
In this basis any $A=A(p)$ can be written as
$$
 A= \lambda \Id + \left(\begin{array}{cc}
                    B & C \\
                    C^\dagger & E
                  \end{array}
\right)\;, \quad \mbox{ while } \
 A_0= \lambda_0 \Id + \left(\begin{array}{cc}
                    0 & 0 \\
                    0 & E_0
                  \end{array}
\right)\;,
$$
where $B$ is a $k\times k$ matrix, with $\lambda=\lambda(p)$, and
with $B$, $C$, and $E$ being analytic functions of $A$, hence smooth
in $p$ (analytic if $A$ is). Since $\dim \Ker(A_0-\lambda_0 \Id)=k$
we have $\det E_0\ne 0$, hence
there exists a neighborhood %$\mcW_p$
of $A_0$ on which $\det E\ne 0$. For $p$
within this neighborhood set $X_i=e_i+\hat X_i$, %$Y=e_2+\hat Y$,
where the vectors $\hat X_i %$, $\hat Y
\in \Vect\{e_{k+1},\ldots,e_N\}$ are given by
$$
\hat X_1 = - E^{-1}C^\dagger e_1\;,\ \ldots\;, \ \hat X_k = -
E^{-1}C^\dagger e_k\;.
$$
Clearly the $X_i$'s %and $Y$
are analytic functions of $A$, thus smooth  (analytic if $A$ is) in
$p$. As $\Ker(A-\lambda \Id)$ has dimension precisely $k$ throughout
$\mcU$ by hypothesis, it easily follows
%, passing to a subset of $\mcW_p$ if necessary,
that the $X_i$'s %and $Y$
span $\Ker(A-\lambda \Id)$.
 \qed

%
%A similar, but simpler, argument shows that each remaining root
%$\mu$ of $W$ is an analytic function of the coefficients $\alpha_i$
%on the set
%%
%$$
% \mcV_1:=\{\alpha_i:\ W(\mu)=0\;,\ W'(\mu)\ne 0\ \mbox{ for some } \
% \mu\in \C\}\subset \C^N
% \;,
%$$
%%
%and that the eigenspaces vary smoothly over sets where each
%eigenvalue has multiplicity one.
%
%It should be clear that the above arguments generalize to
%eigenvalues with higher multiplicities, but this is irrelevant for
%our work here.

\section{Rescalings, $\rho$ and smooth extendibility of $\tell$}
\label{SPaul}

Throughout this appendix the symbol $\ell$ denotes a vector field
satisfying $\nabla_\ell\ell =0$ together with \eq{algspecSpt}. We
assume that the Ricci tensor of $(\mcM,g)$ satisfies the conditions
spelled out in the last part of Theorem~\ref{Tm1v0}. The aim here is
to prove the following:

\begin{Theorem}
\label{Textl}
Suppose that $\ell$ is smooth on  the  intersection $\mcU\cap \mcM$
of a neighborhood $\mcU$ of $ \Scrip $ with $\mcM$,   and let
\beaa  \mcV&:=&\{p\in \Scrip | \ \mbox{$p$ is an end-point of an
integral curve $\gamma$ of $\ell$}\}\subset \Scrip
 \;,
 \\\mcV_{\rho\not\equiv 0}&:=&\{p\in \Scrip | \ \mbox{$p$ is an end-point of an
integral curve $\gamma$ of $\ell$}
 \\
 && \mbox{ \ with $\rho \not \equiv 0$ on
$\gamma$ }\} \subset \mcV
 \;,
 \\
 \mcU_{\rho\not\equiv 0}&:=&\{p\in \Scrip | \ \mbox{$p$ is an end-point of \underline{precisely
 one}
integral curve $\gamma$ of $\ell$}
 \\
 &&
 \mbox{ \ with $\rho \not \equiv 0$ on
$\gamma$}\} \subset \mcV_{\rho\not\equiv 0}
 \;.
\eeaa
Then
\begin{enumerate}
 \item \label{Pextl1}
 The field $\Omega^{-2}\ell$  extends
 smoothly and transversally to a neighborhood of $p\in\mcV$ if and only if $p\in \mcU_{\rho\not\equiv 0}$.
 \item \label{Pextl2}
 The sets
 $\mcV_{\rho\not\equiv 0}$ and $\mcU_{\rho\not\equiv 0}$ coincide, and are open
subsets of $\Scrip $ (perhaps empty).
\end{enumerate}
\end{Theorem}

\proof
% \noindent{\sc Proof of Theorem~\ref{Textl}:}
Point~\ref{Pextl1}: The necessity follows from
Proposition~\ref{Pnotwist}, the sufficiency from
Proposition~\ref{Preg} below. Point \ref{Pextl2} follows from
Proposition~\ref{Preg}.
 \qed

An example of a set $\mcV$ which is the union of  precisely one
generator of $\scrip$ and one generator of $\scrim$ (and is
therefore closed, without interior) is provided by the congruence of
null geodesics with tangent vector $\partial_t+\partial_z$ in
Minkowski space-time. Note that in this example  $\ell$ extends to a
smooth vector field \emph{everywhere} tangent to $\scri$ ,
  and thus $\Omega^{-2} \ell$   extends neither to $\scrip$ nor to $\scrim$.

An example of $\mcV$ which is \emph{not} closed is provided by the
Robinson congruence in Minkowski space-time~\cite[Volume~I,
p.~59]{PenroseRindlerv12}, where $\mcV$ equals $\scri$ with one
generator removed from each of $\scrip$ and $\scrim$.

Theorem~\ref{Textl} has the following corollary:

 \begin{Corollary}
 \label{Cext1}% Let $(\mcM,g)$ be asymptotically simple
Let $\ell$ be smooth on  the  intersection $\mcU\cap \mcM$ of a
neighborhood $\mcU$ of $ \Scrip $ with $\mcM$, and suppose that all
future directed integral curves of $\ell$ in $\mcU$ have end points
on $\scrip$. Then the following conditions are equivalent
 \begin{enumerate}
 \item \label{Cex11}
 $\tell:=\Omega^{-2}\ell$ extends smoothly and transversally
  to $\scrip$.
 \item \label{Cex13}
 $\tilde \rho$ is bounded on  $\mcU$.
 \item \label{Cex12}
 $\rho$ is nowhere vanishing on $\mcU\cap \mcM$.
 \end{enumerate}
 \end{Corollary}

\noindent{\sc Proof:} The implication \emph{\ref{Cex11}}
$\Longrightarrow$ \emph{\ref{Cex13}} is obvious. Next, \eq{sc3}
below shows that $\rho$ does not vanish near $\scrip$ under the
hypothesis of  point \emph{\ref{Cex13}}, but then $\rho$ is nowhere
vanishing by \eq{SC1} as long as the congruence remains smooth, and
the implication \emph{\ref{Cex13}} $\Longrightarrow$
\emph{\ref{Cex12}} follows. Finally, the extendibility part of
\emph{\ref{Cex12}} $\Longrightarrow$ \emph{\ref{Cex11}} follows from
Theorem~\ref{Textl}; transversality follows from the construction in
that Theorem.
 \qed

Before passing to the statement, and proof, of
Proposition~\ref{Preg}, we analyse the transformation properties of
the objects at hand under conformal rescalings.
From the general theory of algebraically-special metrics
\cite{Mason:AlgSpec,Exactsolutions2,LindNewman}, which has been
reviewed in Section~\ref{Smfasvs}, there is a normalized spinor dyad
$(o^A,\iota^A)$ related to the affinely-parameterized vector field
$\ell$ by $\ell^a=o^A\overline{o}^{A'}$, and with the following
restrictions on the spin-coefficients:
\[\kappa=\epsilon=\sigma=\tau=\pi=\lambda=0.\]
In any region in which the complex expansion $\rho$ is non-zero, the
$r$-dependence  of the non-zero spin-coefficients for vacuum,
where $r$ is an affine-parameter along $\ell$ so that $\ell(r)=1$,
has been explicitly found above as:
\bea
\rho&=&-(r+r^0+i\Sigma)^{-1}\label{SC1}\\
\alpha&=&-\alpha^0\rho\label{SC2}\\
\beta&=&-\beta^0\overline{\rho}\label{SC3}\\
\gamma&=&\gamma^0+\frac{1}{2}\rho^2\psi_2^0\label{SC4}\\
\mu&=&\mu^0\overline{\rho}+\frac{1}{2}\rho(\overline{\rho}+\rho)\psi^0_2\label{SC5}\\
\nu&=&\nu^0+\psi^0_3\rho+\frac{1}{2}\psi^1_3\rho^2+\frac{1}{3}\psi^2_3\rho^3\label{SC6}
 \eea
In (\ref{SC1})-(\ref{SC6}), the superscript zero indicates a
function constant along $\ell$ and $\psi^1_{3}$, $\psi^2_{3}$ are
also constant along $\ell$.
%We shall give more details of these
%expressions in the next section.

For the non-vacuum case, $\gamma$, $\mu$ and $\nu$ are given instead
by (\ref{ng3}), (\ref{ng5}) and  (\ref{ng6}), which will be
sufficient for our conclusion below.

With the conformal rescaling $\tg=\Omega^2g$ we obtain
(\cite{Stewart})
\beq \label{res1}
\tnabla_a\tnabla_b\Omega-\frac{1}{2}\Omega^{-1}\tg_{ab}(\tg^{ef}
\partial_e\Omega\partial_f\Omega)=\Omega(-\tPhi_{ab}+\tg_{ab}\tLambda)\;,
\eeq
where, following the usual NP conventions,
\[\Phi_{ab}=-\frac{1}{2}R_{ab}+\frac{1}{8}Rg_{ab}\;\, \quad
\Lambda=\frac{1}{24}R\;,
\]
in terms of the Ricci tensor $R_{ab}$ and scalar curvature $R$, and
the  tilde indicates that these quantities are calculated for $\tg$.

Now $\ell$ is geodesic, shear-free and affinely parameterized for
$g$, and  one readily finds that $\tell=\Omega^{-2}\ell$ has the
same properties for $\tg$. Suppose an affine parameter for $\tell$
is $\trr$, so that $\tell(\trr)=1$, as well as $\ell(r)=1$.
%, so that
%%
%\bel{transr} \frac {d \trr}{dr} = -\Omega^2
% \;.
%\ee
%%
Then $\tell$ is bounded in ${}\tMM$, being a solution of the
equation $\tilde \nabla _\tell \tell =0$ with smooth data at
$\Omega=\epsilon>0$.
Contract (\ref{res1}) with $\tell^a\tell^b$ to find
\beq \label{res2}
\frac{d^2\Omega}{d\trr^2}=\Omega(-\tPhi_{ab}\tell^a\tell^b).\eeq
Integrate this twice along a geodesic of the congruence, fixing the
origin of $\trr$ to be at $\scrip $ (note that $\trr \le 0$ then),
to obtain:
\beal{a60}
 \frac {d\Omega}{d\trr}
 &=& A+
  \int_\trr^0
 \Omega(s)(\tilde \Phi_{ab}\tell^a\tell^b)(s)ds \;,
 \\
 \Omega&
 =&A\trr+ \int_\trr^0  (\trr -s)
 \Omega(s)(\tilde \Phi_{ab}\tell^a\tell^b)(s)ds \;,
 \eeal{a61}
where $A$ is a constant of integration which can be written as
\[
A=\frac{d\Omega}{d\trr}\Big|_{\scrip }=\tell^a\Omega_{,a}|_{\scrip
}\;.
\]
(The limit is negative since $\Omega$ decreases towards $\scrip$).

 Suppose that $p\in \scrip$ is an end-point of an integral curve of
$\ell$. Then   $\tell$ is transverse to $\scrip $ at $p$ and we
conclude that $A$ is nonzero there. We have a remaining freedom to
multiply $\ell$ and hence also $\tell$ by a positive function and we
may use this to set $A=-1$.  Now
\[1=\tell(\trr)=\Omega^{-2}\ell(\trr)=\Omega^{-2}\frac{d\trr}{dr}\;,\]
from which
\bel{abovesc2}
 r\Omega\to1\quad\mathrm{as}\quad r\to\infty
 \;,
 \quad \mbox{and} \quad      \frac{d\trr}{dr}=\Omega^{2}
 \;.
\ee

The chosen rescaling of $\ell$ implies the following rescalings for
the null tetrad
\[\tell^a=\Omega^{-2}\ell^a\;,\quad \tm^a=\Omega^{-1}m^a\;,\quad
\tn^a=n^a\;,
\]
the following for the corresponding one-forms:
\[\tell_a=\ell_a\;,\quad \tm_a=\Omega m_a\;,\quad \tn_a=\Omega^2n_a\;,\]
and the following for the spinor dyad:
\be
\tom^A=\Omega^{-1}o^A\;,\quad \tiota^A=\iota^A\;,\quad \tom_A=o_A\;,\quad \tiota_A=\Omega\iota_A\;,
\label{resc}
\ee
while the spin-coefficients change according to:
\bea
\talpha&=&\Omega^{-1}\alpha-\Omega^{-2}\;\overline{\delta}\Omega
 \;,\label{sc1}\\
\tbeta&=&\Omega^{-1}\beta
 \;,\label{sc2}\\
\trho&=&\Omega^{-2}(\rho-\Omega^{-1}D\Omega)
 \;,\label{sc3}\\
\ttau&=&-\Omega^{-2}\delta\Omega
 \;,\label{sc4}\\
\tgamma&=&\gamma-\Omega^{-1}\Delta\Omega
 \;,\label{sc5}\\
\tpi&=&\Omega^{-2}\;\overline{\delta}\Omega
 \;,\label{sc6}\\
\tmu&=&\mu-\Omega^{-1}\Delta\Omega
 \;,\label{sc7}\\
\tnu&=&\Omega\nu
 \;,\label{sc8} \eea
as well as $\tilde \epsilon=\tilde \kappa=\tilde \sigma=\tilde\lambda=0$. From
(\ref{res1})
\[\tD\tdelta\Omega-(\tD\tm^a)\partial_a\Omega=-\Omega\tPhi_{ab}\tell^a\tm^b:=-\tPhi_{01}\Omega\;,\]
or with (\ref{sc6}) and the definition of $\tpi$ and $\tdelta$:
\bel{befrhobis} \tD(\Omega^{-2}\delta\Omega)=-\tPhi_{01}
 \;.
\ee
%\ee]
%

We are ready to prove now

\begin{Proposition}
\label{Preg} The set $\mcV_{\rho\not\equiv 0}$ is open and coincides
with $\mcU_{\rho\not\equiv 0}$. Moreover
 the field  $\Omega^{-2}\ell$ extends by continuity to a
smooth vector field $\tilde \ell$ on $\mcV_{\rho\not\equiv 0}$.
\end{Proposition}

\proof
Consider an integral curve $\Gamma$ of $\ell$ which has an end point
on $\scrip$ at $p\in \mcV_{\rho\not\equiv 0}$. Then $\Gamma$ can be
extended to a null geodesic with tangent $\tell$, still denoted by
$\Gamma$, which meets $\scrip$ transversally at $p$. There exists
$\epsilon_0>0$ so that $\Gamma$ meets all the level set $
\{\Omega=\epsilon\}$, $0\le \epsilon\le \epsilon_0$ transversally.
Let $\mcW\subset \{\Omega=\epsilon_0\}$ be a small conditionally
compact open neighborhood of $\Gamma\cap \{\Omega=\epsilon_0\}$,
%with coordinates $w^A$,
on which $\rho$
is not vanishing, and to which $\ell$ is transverse. %Because the
%claim is purely local, in the argument below we   replace (without
%loss of generality) $\tmcM$ by a conditionally compact neighborhood
Let the set
$$\tmcO \subset \tmcM
$$
be the union of points obtained by flowing $\mcW$ along the
geodesics tangent to $\tell$ from
$\mcW$ to $\scrip$. %, sufficiently small so that $\ell$ is transverse
%to $\{\Omega=\epsilon_0\}$ in the new space-time $\tmcM$.
 We let
$$
\mcO=\tmcO \cap \mcM
$$
denote the intersection of $\tmcO$ with the original space-time
$\mcM$.

 We start by showing that the tilded
spin coefficients are uniformly bounded on $\mcO$. To see that,
integrate \eq{befrhobis} to find that $\Omega^{-2}\delta\Omega$ is
bounded up to  $\scrip $, and therefore, by (\ref{sc4}) and
(\ref{sc6}), so are $\ttau$ and $\tpi$. From (\ref{SC1})-(\ref{SC3})
and (\ref{sc1})-(\ref{sc3}), we may conclude boundedness of
$\talpha$, $\tbeta$. For $\trho$, straightforward manipulations
using \eq{a61} lead to the following form of \eq{sc3}:
\bean
 \trho
 &=&
  \frac{1 }{\Omega(r+i\Sigma)}\Big[i\Sigma- \frac {1+r\trr}{\Omega}
  + \frac  {r}{\Omega} \int_\trr ^0(\trr- s )
   {\Omega(s)}  (\tilde
\Phi_{ab}\tell^a\tell^b)(s)ds \Big]
 \\
 &&
 -  \frac {1}{\Omega} \int_\trr^0
  {\Omega(s)}  (\tilde \Phi_{ab}\tell^a\tell^b)(s)ds
 \;.
 \eeal{rhobis0}
Note that $\Omega r\to 1$ and  $r\trr \to -1$ as $\trr$ approaches
zero, and boundedness of each term in \eq{rhobis0} easily follows.
 For $\tgamma$, we return to (\ref{res1}) and contract with
$\tell^a\tn^b$ to find
\bel{DDel}
 \tD(\Omega^{-1}\Delta\Omega)=|\Omega^{-2}\delta\Omega|^2-\tPhi_{11}+\tLambda
 \;,
\ee
% \]%
using what we already have. Integrate this to find that
$\Omega^{-1}\Delta\Omega$ is bounded at $\scrip $ and therefore so
also is $\tgamma$, from (\ref{SC4}). Finally, from (\ref{SC5}),
\ref{SC6}), (\ref{sc7}) and (\ref{sc8}), $\tmu$ and $\tnu$ are
bounded.

In the non-vacuum case, we need the modified expressions
(\ref{ng3}), (\ref{ng5}) and (\ref{ng6}) for $\gamma$, $\mu$ and
$\nu$ but the conclusion is the same.

Now
\begin{eqnarray}
\nonumber
\tnabla_a\tell{}^b&=&\tell_a((\tgamma+\overline{\tgamma})\tell{}^b-\ttau\overline{\tm}{}{}^b-\overline{\ttau}\tm{}^b)
 \\
 \nonumber
&&-\tm_a((\talpha+\overline{\tbeta})\tell{}^b-\trho\overline{\tm}{}^b-\overline{\trho}\tm{}^b)
 \\
&&-\overline{\tm}_a((\overline{\talpha}+\tbeta)\tell{}^b-\overline{\trho}\tm{}^b-\trho\overline{\tm}{}^b)\;,
 \label{dervs}
\end{eqnarray}
with similar expressions for the derivatives of $\tn$ and $\tm$, and
we have shown that all the covariant derivatives of the tetrad are
uniformly  bounded. It follows that the tetrad $\tell$, $\tm$,
$\tn$, is uniformly Lipschitz, and therefore extends to a Lipschitz
continuous tetrad on the $\tmcM$-closure $\bmcO\supset\tmcO$ of
$\mcO$. In particular the extended vector field $\tell$ is Lipschitz
continuous. This implies that the map obtained by flowing along the
geodesic with initial tangent  $\tell$  from $\scrip$ for an affine
parameter distance $\trr$ defines a Lipschitz continuous function of
the coordinates, say $v^A$, on $\scrip$: indeed, by definition we
have, in any smooth coordinate system near $\scrip$,
\bel{xcontr}
  x^\mu(\trr,v^A)- x^\mu(\trr,v'^A)= -\int_\trr^0\Big(
\tell^\mu(x^\nu(s,v^A))- \tell^\mu(x^\nu(s,v'^A)\Big) ds
 \;,
\ee
%$$
%
and the Lipschitz character of $v^A\to x^\mu(\trr,v^A) $ follows
from the Gronwall inequality.

We now show (uniform) Lipschitz continuity of the connection
coefficients. First,  from \eq{a60}-\eq{a61}, the functions
$\Omega$, $\Omega/\trr$ and $d\Omega/d\trr$ are now  uniformly
Lipschitz in the variables $(\trr, v^A)$ by a calculation similar to
that in \eq{xcontr}. Next, we want to show Lipschitz continuity of
the right-hand-side of \eq{rhobis0}, which we rewrite in the
following way, convenient for the purposes here:
\bean
 \trho
 &=&
  \frac{1 }{\Omega(r+i\Sigma)}\Big[i\Sigma- \frac {1+r\trr}{\Omega}+r\trr\frac  {\trr}{\Omega(\trr)} \int_\trr ^0 \Big(1-  \frac s \trr \Big)
\frac  {\Omega(s)}{s} \frac s \trr (\tilde
\Phi_{ab}\tell^a\tell^b)(s)ds \Big]
 \\
 &&
 -  \frac {\trr}{\Omega(\trr)} \int_\trr^0
 \frac {\Omega(s)}{s}\frac s \trr (\tilde \Phi_{ab}\tell^a\tell^b)(s)ds
 \;.
 \eeal{rhobis}
%. From
Consider the function
$$
 h:= r\trr +1
 \;;
$$
it follows from \eq{abovesc2} that $h$ satisfies the equation
$$
\trr \frac {dh}{d\trr} = h +H\;, \quad H = \Big(\frac
{\trr}\Omega\Big)^2-1
 \;,
$$
where the function $H$ is already known to be uniformly Lipschitz in
$v^A$. Integration gives
\bel{eq}
 h = C\tilde r\Big(1 + \int _{\trr_0}^\trr \frac{H(s)}{s^2} ds\Big)
 \;,
\ee
and uniform Lipschitz continuity of $h$ --- and hence also of
$r\trr$
--- follows by straightforward estimations. But now
$$
 \Omega r = \Big(\frac \Omega \trr\Big) (  r \trr)
$$
is uniformly Lipschitz  as well.

Rewriting \eq{a61} with $A=-1$ as
$$
 \frac \Omega {\trr} + 1
 = \int_\trr^0  (1 -\frac s\trr) \frac{
 \Omega(s)} s s (\tilde \Phi_{ab}\tell^a\tell^b)(s)ds \;,
$$
we find that $\Omega/ \trr+1$ is $O(\trr^2)$, with $v^A$-H\"older
modulus of continuity also being $O(\trr^2)$. But then
$$
H = \frac { \Big( 1 - \frac \Omega \trr\Big)\Big( 1 + \frac \Omega
\trr\Big)} {\Big(   \frac \Omega \trr\Big)^2}
$$
is $O(\trr^2)$, with $v^A$-H\"older modulus of continuity
$O(\trr^2)$. Rewriting \eq{eq}  as
\bel{eq2}
 \frac h \trr = C \Big(1 + \int _{\trr_0}^\trr \frac{H(s)}{s^2} ds\Big)
 \;,
\ee
we conclude that $h/\trr$ is uniformly Lipschitz continuous in
$v^A$. It follows that $h/\Omega= (h/ \trr)(\trr /\Omega)$ also is.
From the right-hand-side of \eq{rhobis} we  conclude that $\trho$ is
uniformly Lipschitz  in $v^A$.

To continue, integration of \eq{befrhobis} shows that
$\Omega^{-2}\delta\Omega$ is a Lipschitz function of $v^A$, which in
turn justifies Lipschitz continuity of $\ttau$ and $\tpi$.
Furthermore, the uniformly Lipschitz character of the flow of
$\tell$ implies that all the functions such as $\Sigma$, $\alpha^0$,
\emph{etc.}, are Lipschitz continuous functions of $v^A$, hence ---
by composition --- Lipschitz continuous functions on $\tmcM $. This,
together with \eq{SC6} and \eq{sc8} immediately shows that $\tnu$ is
Lipschitz continuous.  From what has been said and from
\eq{SC1}--\eq{SC3}, \eq{sc1}--\eq{sc2}   we conclude that $\talpha$
and $\tbeta$ are uniformly Lipschitz continuous. Finally,
integration of the right-hand-side of \eq{DDel} gives uniform
Lipschitz continuity of $\Omega^{-1}\Delta\Omega$ and hence, in view
of  \eq{SC4}, \eq{SC5}, \eq{sc5} and \eq{sc7}, that of $\tgamma$ and
$\tmu$.

But now the right-hand-side of \eq{dervs} is uniformly Lipschitz
continuous, and hence $\tnabla \tell$ extends to a $C^{1,1}$ vector
field on $\bmcO$. Similarly the remaining elements of the tetrad are
$C^{1,1}$   on $\bmcO $.

The vector field $\tell$ is transverse to $\scrip$ at $p$ by
hypothesis,   further $\tell$  is transverse to $\mcW$, and the
implicit function theorem applied to the map obtained by flowing
from $\mcW$ to $\scrip$ along $\tell$ provides  a diffeomorphism
from a neighborhood of $\Gamma\cap \mcW$ to a neighborhood of $p$
within $\scrip$. This shows in particular that $\mcV_{\rho\not\equiv
0}$ contains a neighborhood of $p$, hence $\mcV_{\rho\not\equiv 0}$
is open. Further,  every point near $p$ is the end point of a unique
element of the congruence generated by $\ell$, so that
$\mcV_{\rho\not\equiv 0}=\mcU_{\rho\not\equiv 0}$ near $p$.

One can iterate the regularity argument above as many times as the
differentiability of the metric allows, obtaining each time one more
degree of differentiability of $\tell$ which, for smooth conformal
boundary extensions, proves smoothness of $\tell$ near $p$.

Since $p\in \mcV_{\rho\not\equiv 0}$ is arbitrary,
Proposition~\ref{Preg} follows.
 \qed

\bibliographystyle{amsplain}
%\bibliographystyle{/usr/share/texmf/tex/revtex/prsty}
%\bibliography{%ptjjk,
%../../references/newbiblio,%
%../../references/reffile,%
%../../references/bibl,%
%../../references/Energy,%
%../../references/hip_bib,%
%../../references/netbiblio}
\bibliography{%ptjjk,
../references/newbiblio,%
../references/newbib,%
../references/reffile,%
../references/bibl,%
../references/Energy,%
../references/hip_bib,%
../references/netbiblio,../references/addon}
%\texttt{\input{READMEl}}

\end{document}